\documentclass[fleqn,usenatbib]{emulatemnras}

\def\MyMNRAS#1{#1}

\usepackage{graphicx}
\usepackage{dcolumn}
\usepackage{bm}
\usepackage{epsf}
\usepackage{amsmath}
\usepackage{amssymb}
\usepackage{amsfonts}
\usepackage{epstopdf}
\usepackage[percent]{overpic}
\usepackage[normalem]{ulem}
\usepackage{footnote}
\usepackage{dblfloatfix}
\usepackage{tikz}

\makesavenoteenv{tabular}

\MyMNRAS{\usepackage{pdflscape}}

\def\figeps#1{./Figures/#1-eps-converted-to.pdf} 

\def\fig#1{./Figures/#1}
\def\myfig#1{#1}

\newcommand{\muS}{{\mu_{S}}}

\MyMNRAS{\bibliographystyle{mnras}}

\newcommand{\MyPlace}{h!}

\newcommand{\ie}{\emph{i.e.} }
\newcommand{\eg}{\emph{e.g.,} }

\newcommand{\be}{\begin{equation}}
\newcommand{\ee}{\end{equation}}
\newcommand{\bea}{\begin{equation*}}
\newcommand{\eea}{\end{equation*}}
\newcommand{\beqr}{\begin{eqnarray} \nonumber}
\newcommand{\eeqr}{\end{eqnarray}}
\newcommand{\beqrb}{\begin{eqnarray}}
\newcommand{\eeqrb}{\nonumber \end{eqnarray}}
\newcommand{\fin}{\mbox{ .}}
\newcommand{\coma}{\mbox{ ,}}

\newcommand{\gr}{\mbox{ g}}
\newcommand{\cm}{\mbox{ cm}}

\newcommand{\se}{\mbox{ s}}

\newcommand{\Gyr}{\mbox{ Gyr}}

\newcommand{\km}{\mbox{ km}}

\newcommand{\kpc}{\mbox{ kpc}}
\newcommand{\Mpc}{\mbox{ Mpc}}

\newcommand{\keV}{\mbox{ keV}}

\newcommand{\K}{\mbox{ K}}

\newcommand{\const}{\mbox{const.}}

\newcommand{\grad}{\bm{\nabla}}

\newcommand{\unit}[1]{\bm{\hat{#1}}}

\usepackage{xcolor}
\definecolor{darkgreen}{rgb}{0.0,0.5,0.0}

\newcommand{\MyTitle}{{Quasi-spiral solution to the mixed intracluster medium and the universal entropy profile of galaxy clusters
}}
\newcommand{\MySTitle}{{Quasi-spiral ICM}}

\MyMNRAS{
    \title[\MySTitle]{\MyTitle}
	\author[Keshet, Raveh, \& Ghosh]{Uri Keshet\thanks{E-mail: ukeshet@bgu.ac.il}, Itay Raveh, \& Arka Ghosh\\
	Physics Department, Ben-Gurion University of the Negev, POB 653, Be'er-Sheva 84105, Israel}
}

\MyMNRAS{
    
\begin{document}
    \pubyear{2023}
    \label{firstpage}
    \pagerange{\pageref{firstpage}--\pageref{lastpage}}
    \maketitle
}

\begin{abstract}
Well-resolved galaxy clusters often show a large-scale quasi-spiral structure in deprojected density $\rho$ and temperature $T$ fields, delineated by a tangential discontinuity known as a cold front, superimposed on a universal radial entropy profile with a linear $K(r)\propto T\rho^{-2/3}\propto r$ adiabat.
We show that a spiral structure provides a natural quasi-stationary solution for the mixed intracluster medium (ICM), introducing a modest pressure spiral that confines the locally buoyant or heavy plasma phases.
The solution persists in the presence of uniform or differential rotation, and can accommodate both an inflow and an outflow.
Hydrodynamic adiabatic simulations with perturbations that deposit angular momentum and mix the plasma thus asymptote to a self-similar spiral structure.
We find similar spirals in Eulerian and Lagrangian simulations of 2D and 3D, merger and offset, clusters.
The discontinuity surface is given in spherical coordinates $\{r,\theta,\phi\}$ by $\phi(r,\theta)\propto \Phi(r)$, where $\Phi$ is the gravitational potential, combining a trailing spiral in the equatorial ($\theta=\pi/2$) plane and semicircles perpendicular to the plane, in resemblance of a snail shell.
A local convective instability can develop between spiral windings, driving a modified global instability in sublinear $K(r)$ regions; evolved spirals thus imprint the observed $K\propto r$ onto the ICM even after they dissipate.
The spiral structure brings hot and cold phases to close proximity, suggesting that the observed fast outflows could sustain the structure even in the presence of radiative cooling.
\end{abstract}

\date{Accepted ---. Received ---; in original ---}

\MyMNRAS{
\begin{keywords}
galaxies: clusters: general - galaxies: clusters: intracluster medium - hydrodynamics - X-rays: galaxies: clusters
\end{keywords}
}

\section{Introduction}
\label{sec:Introduction}

X-ray imaging uncovered an abundance of large-scale quasi-spiral thermal structures in the intracluster medium (ICM) of well-observed galaxy clusters \citep[\eg][and references therein]{MarkevitchEtAl03Proc,ClarkeEtAl04,Keshet2012, UedaEtAl2020}.
Such a spiral structure can extend to a distance $r$ of a few $100\kpc$ from the centre of the cluster, and is delineated by piecewise spiral X-ray edges known as cold fronts \citep[CFs;][]{ClarkeEtAl04, TanakaEtAl06, MarkevitchVikhlinin07}.
ICM spiral structures are often interpreted as sloshing oscillations driven by mergers \citep{MarkevitchEtAl01}, possibly involving only a dark matter subhalo \citep{TittleyHenriksen05,AscasibarMarkevitch06}, by weak shocks or acoustic waves displacing cold central plasma \citep{ChurazovEtAl03, FujitaEtAl04}, or by an inspiraling subcluster core \citep{Clarke04}.
However, there is evidence that such spirals persist in otherwise very relaxed clusters, reflect a long-lived spiral composite flow \citep{Keshet2012} that combines a fast outflow and a slow inflow \citep{NaorEtAl20, NaorKeshet20},
and play a role in offsetting the cooling instability of the core; the spiral structure could therefore be sustained by outflows from the central active galaxy, regulating a spiral cooling flow \citep{Keshet2012, Inoue21}.

An interesting and arguably related observation is the universal radial profile of the $s\propto \ln K$ specific entropy in galaxy groups and clusters, typically quantified in terms of the adiabat $K\equiv k_B T n^{1-\Gamma}$, where $n$ and $T$ are the ICM particle number density and temperature, $\Gamma=5/3$ is the adiabatic index, and $k_B$ is the Boltzmann constant.
When properly deprojected, a simple, approximately linear $K(r)\propto r$ profile usually emerges over a wide mass range
with a universal normalization \citep{Pratt05, Piffaretti05, Donahue06, Sanderson09, Panagoulia14, ReissKeshet2015}.
Such a simple profile, oblivious to the temperature peak at the edge of the core and robust to the presence of ongoing cooling, merger, and active galactic nucleus (AGN) activity, must be sustained by some dynamical mechanism \citep{ReissKeshet2015}.
In the presence of a spiral structure, this entropy profile is locally consistent with a balance between radiative cooling and azimuthal heat conduction or radial heat advection, but the spiral structure is likely to play a more direct dynamical role in regulating the entropy profile \citep{ReissKeshet2015}.

A spiral CF is an edge-on projection of a tangential \citep{KeshetEtAl10} discontinuity, revealing a sharp temperature jump (\ie a sharp increase, henceforth) and density drop (\ie decrease) as one crosses outside (\ie with increasing radius) the CF, as required for Rayleigh-Taylor stability \citep[][and references therein]{MarkevitchVikhlinin07}.
Hydrostatic equilibrium is found to be broken along such a CF \citep{MarkevitchEtAl01}, indicating a fast, Mach $\sim0.8$ flow inside the CF \citep{KeshetEtAl10, NaorKeshet20}.
While the temperature jump and density drop are of order $30\%$ to $100\%$ for a typical CF, the thermal pressure shows a more modest, $\sim 10\%$ jump, indicating the presence of shear-amplified magnetic fields inside the discontinuity \citep{ReissKeshet2014, NaorKeshet20}.
The entropy and metallicity profiles along the CF indicate that the fast flow inside the CF is a nearly sonic outflow, whereas the plasma above the CF is a slow, Mach $\sim0.03$ inflow \citep{NaorEtAl20}.
Away from the discontinuity, variations in the thermal properties are more gradual, the fast flow is likely confined to the vicinity of the CF \citep{KeshetEtAl10}, and the overall ICM typically appears relaxed.

Modeling the observed CFs suggests that the deprojected discontinuity surface is given in spherical coordinates $\{r,\theta,\phi\}$ by $r\simeq r_{d}(\phi)f(\theta)$, where $r_{d}$ specifies the spiral discontinuity pattern in some preferred, $\theta=\pi/2$ equatorial plane, and $f(\theta)$ traces the profile perpendicular to this plane at a fixed $\phi$.
Projected results are typically consistent with $f(\theta)\simeq 1$, giving a radius of curvature $R_\theta\simeq r_{d}(\phi)$ in the $r$--$\theta$ plane consistent with semicircles \citep[][and Ghosh et al., in prep.]{NaorEtAl20}, although fast flows and unrelaxed spirals show a larger, $R_\theta>r_{d}$ radius of curvature \citep{Keshet2012, ReissKeshet2015}.

These observations suggest that a quasi-spiral configuration may provide a quasi-steady-state solution for the ICM, remaining stable over many dynamical times \citep{Keshet2012}.
This conclusion is supported by adiabatic, hydrodynamic \citep[\eg][]{AscasibarMarkevitch06, ZuHoneEtAl11, RoedigerEtAl11, ZuHoneEtAl16} and magnetohydrodynamic \citep[MHD;][]{ZuHoneEtAl15, WernerEtAl15} simulations of merger events, found to generate a long-lived spiral pattern at late times, resilient to subsequent minor mergers \citep{VaezzadehEtAl22}.
Although such simulations are typically unrealistic, neglecting the strong radiative cooling and feedback from the central AGN \citep[otherwise the core collapses rapidly; \eg][]{ZuHoneEtAl10}, and fail to reproduce the observed strong shear and fast outflows, merger parameters can be found to match the observed spiral morphology.

We model this putative quasi-steady state, analytically and numerically, in order to address a range of open questions.
For example, are three-dimensional (3D) effects essential, or can a steady-state spiral form in two-dimensions (2D)?
Studies of ICM spirals in 2D were not carried out, to our knowledge, until now.
The pressure profile is approximately radial, so how can the hot and cold spiral phases coexist at a given radius without the buoyant rising of the former or sinking of the latter?
Are radial flows, shear, or angular momentum essential for the survival of the spiral?
What determines the discontinuity profile $r_{d}(\phi)$, in particular the type of spiral and its trailing or leading orientation?
How are the details of the perturbation and of physical processes such as viscosity imprinted on the spiral?
And finally, do spirals in adiabatic simulations drive the entropy towards its universal profile?
A model that resolves these questions, even if it lacks key physical ingredients such as AGN feedback and radiative cooling, would provide a better understanding of observations and could serve as a basis for more realistic models.

The paper is organized as follows.
In \S\ref{sec:Model}, we present our assumptions (\S\ref{subsec:ModelAssumptions}) and the spiral-structure model, first in the self-similar limit (\S\ref{subsec:SelfSimilar}) and then for a more general distribution (\S\ref{subsec:NonSelfSimilar}).
The numerical simulations are presented in \S\ref{sec:Simulations}, with their different set-ups (\S\ref{subsec:SimSetup}), and shown to evolve (\S\ref{subsec:QuasiSteadyState}) into a quasi-steady state (\S\ref{subsec:QuasiSpiral}) that is consistent with the model in terms of spiral development (\S\ref{subsec:SpiralEvolution}), geometry (\S\ref{subsec:MeasuredGeometry}), azimuthal structure (\S\ref{subsec:SpiralAzimuthals}), and radial structure (\S\ref{subsec:SpiralRadials}), in particular imprinting a linear $K(r)\propto r$ profile onto the ICM (\S\ref{subsec:SpiralEntropy}).
The results are summarized and discussed in \S\ref{sec:Discussion}.

We adopt a $\Lambda$CDM model with a Hubble parameter $H_0=70\,\mbox{km}\,\mbox{s}^{-1}\,\mbox{Mpc}^{-1}$,
a matter fraction $\Omega_m=0.3$, a baryon fraction $f_b\equiv\Omega_b/\Omega_m=0.17$ giving a mean mass $\bar{m}\simeq 0.59m_p$, where $m_p$ is the proton mass, and assume a $\Gamma=5/3$ adiabatic index for the plasma.
Radiative cooling and AGN feedback are neglected under the common assumption that feedback somehow stabilizes the cluster against the cooling instability.

\section{Spiral ICM model}
\label{sec:Model}

\subsection{Assumptions and governing equations}
\label{subsec:ModelAssumptions}

\begin{figure}
\begin{center}
\myfig{
\includegraphics[width=1.0\linewidth,trim={1.1cm 3.9cm 1.7cm 3.4cm},clip]{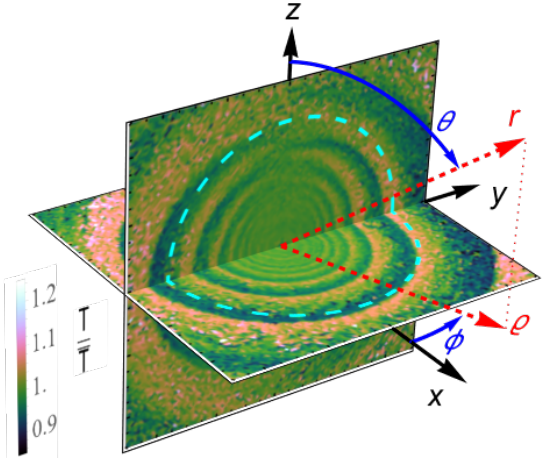}
}
\end{center}
\caption{
Set-up illustration. Geometry and notations (curves and labels) are superimposed on the quasi-steady temperature distribution (normalised to the radial average, colourbar) in our nominal offset 3D-GADGET simulation (at $t=30\Gyr$; planar cut width is $1\Mpc$).
The $x$--$y$ spiral plane shows spiral structure (concentric circles are shown for reference, dashed cyan), whereas the $y$--$z$ perpendicular plane shows approximately concentric semicircles.
\label{fig:setup}
}
\end{figure}

We model the ICM as a viscous but otherwise ideal fluid, neglecting diffusion, radiative cooling, heat conduction, and magnetic fields.
Momentum conservation then reads \citep[\eg][]{LandauLifshitzFluid}
\begin{align}\label{eq:NSFull}
    \rho \frac{dv_i}{dt} & = \rho \left(
        \frac{\partial v_i}{\partial t}+ v_k\frac{\partial v_i}{\partial x_k}
    \right)
    \\
    & = -\frac{\partial P}{\partial x_i} + \rho g_i +\frac{\partial }{\partial x_k}\left[
        \mu \left(
            \frac{\partial v_i}{\partial x_k}+\frac{\partial v_k}{\partial x_i}-\frac{2}{3}\delta_{ik}\frac{\partial v_l}{\partial x_l}
        \right)
    \right]
    \nonumber \coma
\end{align}
where we used the Einstein summation convention and Cartesian coordinates $\{x_i\}_{i=1}^3$ or equivalently $\{x,y,z\}$, with $z$ chosen along $\theta=0$; see Fig.~\ref{fig:setup}.
Here, $\rho$ is the mass density, $\bm{v}$ the velocity, $P$ the pressure, $\bm{g}\equiv -\grad\Phi$ the gravitational acceleration field, $\Phi$ the gravitational potential, and $\mu$ the shear viscosity; bulk viscosity is neglected.
Spiral perturbations in the gravitational field due to the baryonic ICM are small and neglected henceforth, so $\bm{g}\simeq g(r)\unit{r}$ is approximated as static and radial.

The momentum equation is supplemented by the continuity, \ie mass, equation,
\begin{equation} \label{eq:MassCons}
    \frac{d\rho}{dt}=\frac{\partial\rho}{\partial t}+v_k \frac{\partial \rho}{\partial x_k}=0 \coma
\end{equation}
and the energy equation
\begin{equation}\label{eq:EnergyCons}
    \frac{ds}{dt}=\frac{\partial s}{\partial t}+v_k \frac{\partial s}{\partial x_k}=0 \coma
\end{equation}
where $s=c_V\ln(K)$ is the specific entropy and we neglected viscous dissipation.
Here, $c_V=(k_B/\bar{m})/(\Gamma-1)$ is the specific heat at constant volume and $K=\rho^{-\Gamma}P$ is the adiabat.
Under present assumptions, in the absence of diffusion, heating, cooling, and supersonic motion, these two equations become trivial in the co-rotating frame, defined below, so will be of little use for what follows.

Most of the ICM volume is thought to be subsonic, with only weak shear and viscous forces.
Inasmuch as the gravitational field can be approximated as radial and fixed, momentum conservation \eqref{eq:NSFull} then becomes, to leading order,
\begin{equation} \label{eq:NSr}
\grad P(t;\bm{r}) \simeq \rho(t;\bm{r}) g(r) \unit{r} \, ;
\end{equation}
second-order corrections are introduced later.
Our simulations (see \S\ref{sec:Simulations}) show that the spiral ICM structure persists when the flow subsides into approximately uniform rotation, whereby corrections to Eq.~\eqref{eq:NSr} vanish and Eqs.~\eqref{eq:MassCons} and \eqref{eq:EnergyCons} are satisfied identically; the spiral remains imprinted on the ICM even when this rotation slows down further by more than an order of magnitude.
Although the density field is not continuous at the discontinuity, it is finite everywhere, so integrating Eq.~\eqref{eq:NSr} along radial rays yields a pressure field that is continuous, as expected, even at the discontinuity.

Denote $P_0(r)$ and $\rho_0(r)$ as the spherical (\ie without any spiral structure) pressure and density distributions that solve Eq.~\eqref{eq:NSr} for a galaxy cluster, with the prescribed $g(r)<0$.
Consider a putative ICM steady-state that superimposes upon this spherical distribution some spiral thermal structure.
Such a spiral structure should have a preferred axis and a perpendicular, so-called spiral plane, in which the spiral attributes of the distribution are most pronounced.
We choose this preferred axis as the $z$ direction, and the spiral plane as the equatorial, $\theta=\pi/2$ or equivalently $z=0$, plane.
The discontinuity manifold can now be written as $\phi=\phi_d(t;r,\theta)$.
A CF is observed at lines of sight $\bm{l}$ which are tangent to this discontinuity; see \citet{Keshet2012}, \citet{NaorKeshet20}, and Ghosh et al. (in prep.) for discussions of projection effects.

Without loss of generality, let us assume that the spiral opens outwards in the clockwise sense in the $x$--$y$ plane, so $\phi_d$ monotonically decreases with an increasing $r$; see Fig.~\ref{fig:setup}.
Here, crossing the discontinuity with an increasing $r$ is equivalent to crossing it with an increasing $\phi$.
Rayleigh-Taylor stability then requires that $\rho(\phi)$ drops and $T(\phi)$ jumps sharply as $\phi$ increases beyond $\phi_d$.
As $\phi$ is a periodic variable, and observations show monotonic $\rho$ and $T$ behaviours away from the CF, we deduce that $\partial_\phi \rho>0$ and $\partial_\phi T<0$ in most of the volume, except inside the sharp, confined discontinuity transition itself, approximated here as infinitely thin.
Denote the fractional density contrast across the discontinuity as $q\equiv \rho_i/\rho_o>1$, where index $i$ (index $o$) refers to plasma just inside (outside), \ie below (above) the CF.
Although $q$ likely varies along the discontinuity \citep{ReissKeshet2015}, these variations are observed to be small \citep{NaorKeshet20}.

\subsection{Self-similar spiral structure}
\label{subsec:SelfSimilar}

It is instructive to first consider a self-similar spiral structure, which appears to capture the main properties of simulated and observed ICM configurations.

\subsubsection{Self-similar ansatz}
\label{subsubsec:SelfSimilarAnsatz}

Denote the positive azimuthal distance from the discontinuity as
\begin{equation}
0\leq \delta\phi(t;\bm{r})\equiv \phi-\phi_d(t;r,\theta) \leq 2\pi\, ,
\end{equation}
for any point in spacetime.
A spiral structure can be introduced by modifying each thermal state function $A(t;\bm{r})\in\{P,\rho,v_i,T,s,n,K,\ldots\}$, from its spherical steady-state distribution $A_0(r)$, into a function $A=A(t;r,\theta,\delta\phi)$ with some simple dependence upon $\delta\phi$.
In our simplest self-similar ansatz, we approximate the spiral structure, $A(t;r,\theta,\delta\phi)/A_0(r)$, as a function of $\delta\phi$ alone, drastically compactifying the full 3+1 dimensional spacetime dependence once $\phi_d$ is determined.
More precisely, the fractional deviation of $A$ from $A_0$ is approximated as a dimensionless function $\delta A$ of $\delta\phi$, such that
\begin{equation}\label{eq:SSSAnsatz}
A(t;\bm{r}) \simeq \left\{1+ \delta A[\delta\phi(t;\bm{r})]\right\} A_0(r)\fin
\end{equation}

Under the self-similar ansatz \eqref{eq:SSSAnsatz}, hydrostatic equilibrium \eqref{eq:NSr} becomes
\begin{equation} \label{eq:NSr2}
\frac{P_0(r) \partial_r\phi_d(t;r,\theta)}{g(r) \rho_0(r)} \simeq \xi\simeq \frac{\delta P(\delta\phi)-\delta\rho(\delta\phi)}{\delta P'(\delta\phi)}\coma
\end{equation}
decoupling into two separate equations.
One equation determines the geometry of the discontinuity,
\begin{equation}\label{eq:NSrr}
\xi^{-1}\partial_r\phi_d(t;r,\theta) \simeq \frac{g(r)\rho_0(r)}{P_0(r)} = -\Gamma \frac{\Phi_0'(r)}{c_s^2} \coma
\end{equation}
where $c_s$ is the unperturbed sound speed.
The second equation determines the spiral thermal structure,
\begin{equation}\label{eq:NSrY}
\xi\, \delta P'(\delta\phi) \simeq \delta P(\delta\phi)-\delta\rho(\delta\phi) \fin
\end{equation}
The dimensionless coupling constant $\xi$ is positive because $g$ and (for our spiral orientation) $\partial_r\phi_d$ are both negative.
Equation \eqref{eq:NSrr} implies that $\partial_r\phi_d$ is proportional to $\xi$, so this parameter controls the tightness of the spiral: a larger $\xi$ yields a tighter spiral.

As $A(t;r,\theta,\phi_d+\delta\phi)/A_0(r)$ is assumed above to depend only on $\delta \phi$, the fractional contrast of each quantity $A$ across the discontinuity, and in particular the CF density contrast $q$, is constant throughout the discontinuity surface.
Such a simple description of the system is useful, but only approximate; in particular, the contrasts must vanish before reaching the $\theta\in\{0,\pi\}$ poles, otherwise a non-trivial discontinuity would terminate at a point.
A simple generalization of the self-similar spiral \eqref{eq:SSSAnsatz} is to admit also a $\theta$ dependence, such that $\delta A=\delta A(\theta,\delta \phi)$, in which case the fractional contrasts do depend (only) on $\theta$.
Here, Eqs.~\eqref{eq:NSr2}--\eqref{eq:NSrY} remain valid, but $\xi$ may become a function of $\theta$.
Nevertheless, if $\delta P(\theta,\delta\phi)$ is separable, then Eq.~\eqref{eq:NSr2} indicates that $\xi$ remains a constant.
In such a case, and more generally when the $\theta$ dependence of $\xi(\theta)$ is negligible, the following analysis remains qualitatively unchanged.
In particular, the self-similar spiral picture below generalizes $\delta A(\delta\phi)$ to $\delta A(\theta,\delta \phi)$, simply by multiplying $\delta \rho(\delta\phi)$ and $\delta P(\delta\phi)$ by a universal function $\Theta(\theta)$.
Then $q\propto \Theta(\theta)$ approaches unity as $\theta\to\{0,\pi\}$.
For simplicity, in the following we adopt a constant $\Theta(\theta)=1$, postponing a generalization to future work (Ghosh et al., in prep.).

\subsubsection{Discontinuity-surface geometry}
\label{subsec:CDGeometry}

Equation \eqref{eq:NSrr} can be integrated to determine the geometry of the discontinuity, as a function of the unperturbed potential and temperature profiles.
In regions where the temperature changes slowly so $c_s$ is roughly constant, we may then approximate
\begin{equation}
\phi_d(t;r,\theta) \simeq - \frac{\xi\Gamma}{c_s^2} \Phi(r) + f(t;\theta) \fin
\end{equation}
The arbitrary function $f$ can rotate the discontinuity around the $z$ axis as a function of polar angle and time.
However, as $f$ cannot depend on $r$, its effect on an extended spiral pattern is limited.
At small radii, $|\Phi|$ becomes very large, so in a realistic spiral $f$ must be subdominant and not appreciably modify the geometry.
As the radius increases, $f$ could in principle become dominant and its $\theta$-dependence could substantially alter the spiral structure, but this is not seen in simulations.
We may thus approximate $f(t;\theta)\simeq f(t)\equiv \omega(t)t$, contributing only some uniform rotation at a global angular frequency $\omega(t)$.
In a frame co-rotating at angular frequency $\omega(t)$, the azimuthal coordinate of the discontinuity then depends only on $r$,
\begin{equation} \label{eq:SSSpiral}
\phi_{d,{\rm rot}} \equiv \phi_d(t;r,\theta) - \omega(t)t \simeq  - \frac{\xi\Gamma}{c_s^2} \Phi(r) \fin
\end{equation}
This formulation neglects a dependence on $\theta$ and differential rotation, which can be significant before the discontinuity asymptotes to its self-similar state.

As the spiral discontinuity surface $\phi_d$ (in the co-rotating frame, henceforth, dropping the co-rotating frame subscript `rot' for brevity) given by Eq.~\eqref{eq:SSSpiral} is independent of $\theta$, it consists of semicircles perpendicular to the spiral plane, as anticipated above.
Hence, it suffices to determine the spiral pattern in the spiral plane, which, as the equation shows, is proportional to the gravitational potential.
For example, one expects a logarithmic, $\phi_d\propto \log{r}$ spiral in an isothermal sphere distribution,
an approximately hyperbolic, $\phi_d\propto (r+a)^{-1}$ spiral in a \citet{Hernquist90} profile,
and a combined, $\phi_d\propto r^{-1}\log{r}$ spiral in an NFW \citep{NavarroEtAl1997} profile.
The spiral structure obtained from hydrodynamical simulations of a Hernquist profile of scale length $a$ is demonstrated in Fig.~\ref{fig:setup}, and is indeed well-fit by a hyperbolic spiral (see \S\ref{subsec:MeasuredGeometry}).

\subsubsection{Unchanged $\rho(r)$ and $P(r)$ profiles}
\label{subseubsec:UnchangedRhoP}

The radial profiles of thermal quantities, averaged over polar and azimuthal angles, are important, as they are often extracted from observations.
The self-similarity ansatz \eqref{eq:SSSAnsatz} and the $\theta$-independent discontinuity pattern \eqref{eq:SSSpiral} indicate that the radial profile of any thermal state function $A$ is given by
\begin{align}
A(r) & \equiv \frac{1}{4\pi}\int A(\bm{r})d\Omega = \frac{1}{2\pi}\int A_0(r)[1+ \delta A(\delta\phi)] d\phi \nonumber \\
& = \left[1 + \frac{1}{2\pi}\int \delta A(\delta\phi) d\phi \right] A_0(r) = C_A A_0(r) \coma
\end{align}
where $\Omega$ is the solid angle with respect to the centre of the cluster.
Thus, the above self-similar spiral ansatz can modify the radial thermal profiles only by multiplying them by global constants $C_A$, which vary among the different functions $A$.
In practice, the radial profile $A(r)$ may also change due to additional effects, in particular transients found before self-similarity is fully established; indeed, such modifications are very likely for quantities $A$ with a corresponding $C_A\neq 1$.

In the absence of shocks, and neglecting the magnetic layers observed beneath CFs and other nonthermal effects, the thermal pressure is continuous everywhere, including at the tangential discontinuity.
Hence, the integral of Eq.~\eqref{eq:NSrY} along $\phi$ from one side of the discontinuity to the other, which is proportional to $\int \delta P'(\phi)d\phi$, must vanish.
The effect of the spiral on the $P(r)$ profile can thus be related to its effect on the baryon mass $M(r)$ inside $r$,
\begin{equation}\label{eq:IntRho}
\frac{M'(r)}{M_0'(r)}-1 = \frac{1}{2\pi}\int_{\phi_d}^{\phi_d+2\pi} \delta \rho \,d\phi  = \frac{1}{2\pi} \int_{\phi_d}^{\phi_d+2\pi} \delta P \,d\phi \coma
\end{equation}
which is by definition a constant because $\delta \rho$ and $\delta P$ depend on $r$ only through $\delta \phi$.
Hence, when azimuthally averaged, the spiral structure simply multiplies the overall radial profiles of mass, density, and pressure by the same factor, \begin{equation}
\frac{M(r)}{M_0(r)} = \frac{\rho(r)}{\rho_0(r)} = \frac{P(r)}{P_0(r)} = C_\rho = C_P = \const
\end{equation}
Furthermore, assuming that introducing the spiral structure does not modify the total baryon mass it encloses, we find that $C_\rho=C_P=1$, and the integrals in Eq.~\eqref{eq:IntRho} vanish.

We conclude that in the presence of the self-similar spiral, the radial (\ie azimuthally averaged) profiles of density and pressure are the same as in the unperturbed state, and so are the resulting mass and gravitational potential profiles, $M(r)$ and $\Phi(r)$.
However, the radial profiles of temperature, entropy, and other quantities that are not linear combinations of $\rho$ and $P$, are in general modified by the spiral, as shown in \S\ref{subseubsec:ModifiedTandK}.

\subsubsection{Thermal spiral also in pressure}
\label{subseubsec:PressureSpiral}

Equation \eqref{eq:NSrY} indicates that a spiral structure imprinted on the density distribution, \ie $\delta \rho\neq 0$, requires a spiral structure in pressure, too.
While the associated gradients in pressure are typically more subtle than their density and temperature counterparts, and are often overlooked, they are essential for radial force balance.
Namely, after the ICM has been mixed, a pressure spiral is necessary in order to keep the locally dilute or dense plasma from buoyantly rising or sinking.
The pressure force needed to stabilize the spiral is proportional to the pressure gradient, and hence inversely proportional to the distance between spiral windings.
Therefore, while a larger $\xi$ produces a tighter spiral, for a given $q$ it also lowers the amplitude of $|\delta P|$, thus keeping the pressure force approximately unchanged, as shown below.

As mentioned in \S\ref{subsec:ModelAssumptions}, for our choice of spiral orientation, $\partial_\phi \rho>0$ everywhere except within the infinitely thin discontinuity transition.
As long as the spiral density gradients exceed their pressure counterparts, Eq.~\eqref{eq:NSrY} then implies that $\delta P''(\delta\phi)<0$, so the $P(\phi)$ profile is concave.
(This would remain true even if pressure gradients were strong, as long as the ordering $\delta P'(\delta\phi)<-\delta\rho'(\delta\phi)<0$ holds.)
Consequently, as the pressure is continuous, $P(\phi)$ is minimal at the discontinuity $\phi=\phi_d$, increases with $\phi$ for $\phi_d<\phi<\phi_{\rm max}$, and decreases back to its minimum for $\phi_{\rm max} <\phi <\phi_d$.
The resulting $\delta P(\phi)$ profile thus resembles a downward-opening, \ie concave, (possibly distorted) parabola, with its maximum at $\phi_{\rm max}$ close to $\phi_d+\pi$.
An equivalent way to see this is to recall that for $C_\rho=1$, the integrals in Eq.~\eqref{eq:IntRho} vanish, so $\delta\rho$ must be negative just above the discontinuity (where $\delta\phi\simeq 0$), but increases with $\phi$ and is positive just below the discontinuity ($\delta\phi \simeq 2\pi$).
Typically, $|\delta P|<|\delta \rho|$, so Eq.~\eqref{eq:NSrY} shows that a continuous $P(\phi)$ resembles a concave parabola.

\begin{figure}
\begin{center}
\myfig{
\includegraphics[width=1.0\linewidth,trim={0cm 0cm 0cm 0cm},clip]{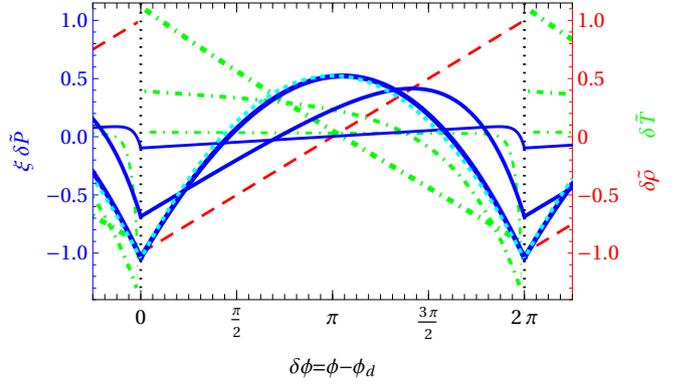}
}
\end{center}
\caption{
Normalised azimuthal profiles of the self-similar spiral $\delta \tilde{A}\equiv (q+1)\delta A /(q-1)$ corrections to density (dashed red; right axis), pressure (solid blue; multiplied by a factor $\xi$ for visibility; left axis), and temperature (green dot-dashed; with $q=3/2$; right axis). Results shown for the linear $\delta\rho$ of Eq.~\eqref{eq:LinearDRho}, with tightness parameters $\xi=10$, $1$, and $0.1$ (thick to thin curves). For a tight spiral, the pressure becomes a concave parabola (the asymptotic limit \eqref{eq:deltaPTight} shown as dotted cyan), and the temperature profile becomes linear. The tangential discontinuity is shown as periodic vertical black dotted lines.
\label{fig:SSParabola}
}
\end{figure}

For concreteness, consider the lowest-order meaningful expansion of $\delta \rho(\delta \phi)$.
The simplest non-trivial profile with CF contrast $q$ is linear, whereby the arguments of \S\ref{subsec:ModelAssumptions} imply that
\begin{equation}\label{eq:LinearDRho}
\delta \rho = \frac{q-1}{q+1}\left(\frac{\delta\phi}{\pi}-1 \right) \fin
\end{equation}
Here, under the self-similar spiral ansatz \eqref{eq:SSSAnsatz}, $q$ is constant throughout the volume, although this can be generalized to $q(\theta)$ as discussed in \S\ref{subsubsec:SelfSimilarAnsatz}.
For such a linear azimuthal density profile, consistent with both observations and simulations, Eq.~\eqref{eq:NSrY} gives the spiral pressure profile
\begin{equation}\label{eq:LinearDP}
\delta P = \frac{q-1}{q+1}\left(\frac{\xi+\delta\phi}{\pi}-1+\frac{2e^{\delta\phi/\xi}}{1-e^{2\pi/\xi}}\right) \coma
\end{equation}
demonstrated in Fig.~\ref{fig:SSParabola}.
The figure shows the normalised profiles
\begin{equation} \label{eq:DeltaTildeA}
\delta\tilde{A}\equiv \frac{q+1}{q-1}\delta A \coma
\end{equation}
for $A\in\{\rho,T,P\}$; for the latter, $\delta\tilde{P}$ is weighted in the figure by $\xi$ so it remains visible in a tight spiral.
In the large $\xi$ limit, this solution asymptotes to the concave parabola
\begin{equation}\label{eq:deltaPTight}
\delta P \simeq \frac{\pi}{6\xi}\frac{q-1}{q+1}\left[1-3\left(1-\frac{\delta\phi}{\pi}\right)^2\right]\coma
\end{equation}
symmetric around $\phi_{\rm max}=\phi_d+\pi$, and diminished in inverse proportion to $\xi$.

Indeed, as shown in \S\ref{sec:Simulations}, pressure spirals are recovered in high-resolution simulations, and the azimuthal profile is found to be approximately given by a concave parabola.
There is evidence for such pressure spirals in observations, too.
In Perseus, a subtle spiral is evident in the projected pressure map \citep{ChurazovEtAl03}, with the maximal pressure seen to be located between the spiral CFs, close to the expected $\phi_{\rm max}\simeq\phi_d+\pi$.
A similar effect was pointed out in the nearly face-on spirals in A2204 and possibly A496 \citep{ReissKeshet2014}.
These observations are based on the electron thermal pressure, which serves as a fair tracer of the total pressure except in the fast flow regions just below the CF \citep[][and references therein]{NaorKeshet20}.

\subsubsection{Modified $T(r)$ and $K(r)$ profiles}
\label{subseubsec:ModifiedTandK}

\begin{figure}
\begin{center}
\myfig{
\includegraphics[width=0.85\linewidth,trim={0.2cm 0cm 1cm 1.0cm},clip]{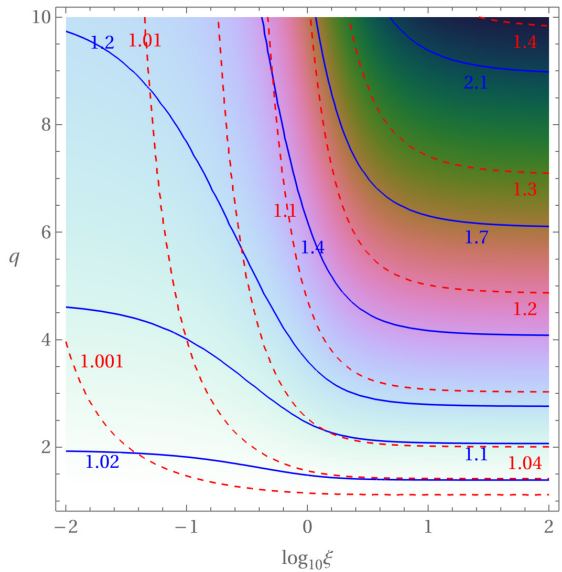}
}
\end{center}
\caption{
Fractional increase in the azimuthally-averaged adiabat, $C_K=K(r)/K_0(r)$ (solid blue contours and colour shading), and temperature, $C_T=T(r)/T_0(r)$ (dotted red contours), due to the self-similar spiral.
\label{fig:SSTandK}
}
\end{figure}

Although incorporating the spiral structure does not alter the radial, \ie azimuthally-averaged profiles of density or pressure, it does modify the profiles of temperature, entropy, and other non-linear combinations of $\rho$ and $P$.
In particular, the $T\propto \rho^{-1}P$ temperature and the $K\propto \rho^{-\Gamma}P$ adiabat in general increase due to the presence of the spiral.
Namely, in the self-similar spiral, where each radial profile is uniformly multiplied by a constant, $T(r)=C_T T_0(r)$ and $K(r)=C_K K_0(r)$, we find that the constants $C_T$ and $C_K$ both generally exceed unity.
In particular, Fig.~ (\ref{fig:SSTandK}) shows the constants $C_K>1$ (solid blue contours) and $C_T>1$ (dashed red contours) obtained for the linear $\delta\rho$ model \eqref{eq:LinearDRho}.
As the figure shows, these constants monotonically increase with the discontinuity contrast $q$ and the spiral tightness $\xi$.
In the $\xi\to\infty$ limit, $C_T=(q+1)\ln(q)/2(q-1)$ and $C_K=(3/2^{8/3})(q+1)^{2/3}(q^{4/3}+q+q^{1/3}+1)/(q^{4/3}+q^{2/3}+q)$.

Although the azimuthally-averaged profiles are simply rescaled by a constant in the self-similar spiral, modifications of the local gradients, not averaged over angles, may destabilize the spiral.
Crossing the discontinuity with an increasing radius, the density drops and the adiabat jumps.
These abrupt changes are compensated away from the discontinuity, where the radial decline in $\rho$ and rise in $K$ become milder with respect to their unperturbed, spherically-symmetric counterparts $\rho_0(r)$ and $K_0(r)$.
Rayleigh-Taylor and convective stability then constrain $q$ and $\xi$, as the radial gradients of $\rho$ and $K$ should remain negative and positive, respectively.

For concreteness, consider the model \eqref{eq:LinearDRho}.
Requiring that $\partial_r \rho<0$ everywhere indicates that a finite contrast limits the spiral tightness,
\begin{equation}
\xi<\frac{2\pi(-\lambda_\rho)c_s^2q}{(q-1)\Gamma v_c^2} \coma
\end{equation}
where $v_c=(-g r)^{1/2}$ is the circular velocity, and we defined $\lambda_\rho$ or more generally
\begin{equation}
\lambda_A\equiv \frac{d\ln A_0}{d\ln r} = \frac{d\ln A(r)}{d\ln r}
\end{equation}
as the radial power-law index of the unperturbed quantity $A_0$ (which is also the slope of the self-similar, azimuthally-averaged $A$, as seen in \S\ref{subseubsec:UnchangedRhoP}).
Equivalently, this result can be written as an upper limit on the contrast,
\begin{equation} \label{eq:LocalRTInstability}
1<q<\left[1-\frac{2\pi(-\lambda_\rho)c_s^2}{\Gamma \xi v_c^2}\right]^{-1} \coma
\end{equation}
except in a loose spiral where $\xi$ is small enough to render the square brackets negative.
Similarly requiring that $\partial_r K>0$ everywhere yields a similar constraint, which to lowest order in $\xi^{-1}$ becomes
\begin{equation} \label{eq:LocalConvectionInstability}
\frac{15\xi}{2\pi} < \frac{9\lambda_K c_s^2}{(q-1)\Gamma v_c^2}-\frac{5q+1}{q+1} \fin
\end{equation}
This constraint again shows that a tight spiral requires a small contrast.

Consider a spiral that is well developed only at small radii, or has a contrast that declines radially.
In such scenarios, and inasmuch as the above self-similar results can be applied locally to parts of such a globally non-self-similar spiral, the radial entropy profile would become shallower, \ie $K(r)$ would increase more slowly.
A similar effect can arise due to inflows or outflows, which introduce high entropy plasma at small radii or low entropy plasma at large radii.
Such effects, which could result in a convective instability but are outside the scope of the self-similar spiral, are discussed in \S\ref{subsec:RadFlowAndInst} below.

\subsection{General spiral structure}
\label{subsec:NonSelfSimilar}

Next, we relax the self-similarity assumption and consider a more general flow pattern.
To separate out a bulk, possibly differential, rotation around the $\unit{z}$ axis, we write the velocity field as
\begin{equation} \label{eq:RotationV}
\bm{v}=\omega(t;r,\theta)\varrho\hat{\bm{\phi}} + \bm{v}_1(t;\bm{r}) \coma
\end{equation}
where the angular frequency $\omega$ is now allowed to vary also spatially, and not only temporally.
Here, we introduced also cylindrical coordinates $\{\varrho,\phi,z\}$, with a cylindrical radius $\varrho\equiv r\sin\theta$; see Fig.~\ref{fig:setup}.
Any variability of $\bm{v}$ in $\phi$ is absorbed in $\bm{v}_1$, with $\omega$ defined such that $\int (\bm{v}_1\cdot\unit\phi)\,d\phi=0$.

If the $\bm{v}_1$ component of the velocity can be neglected, we may isolate the equatorial plane, or any plane parallel to it, and simulate its flow in two dimensions (2D).
Note that even if a symmetry implies that $v_z=0$ in the equatorial plane, the flow there is not 2D if the radial flow within the plane is significant, $v_r\neq0$, as the continuity equation would then couple the plane to the flow outside it \citep{Keshet2012}.
The simulations presented in \S\ref{sec:Simulations} demonstrate that at late times, spiral structures indeed develop in 2D simulations in good agreement with their spiral-plane counterparts in 3D simulations, consistent with $\bm{v}_1$ becoming negligible.

\subsubsection{Advection of the discontinuity}
\label{subsec:AdvectionDiscontinuity}

Fluid elements cannot cross a tangential discontinuity, which is therefore simply advected with the flow.
If $\bm{v}_1$ or at least its temporal average can be neglected, advection thus evolves the discontinuity as
\begin{equation}\label{eq:CFPattern}
\phi_{d}(t_1;r,\theta)=\phi_{d}(t_0;r,\theta)+\int_{t_0}^{t_1} \omega(t;r,\theta)\,dt
\end{equation}
between any two times, $t_0$ and $t_1$.
The spiral typically winds up due to the $r$ dependence of the integral, which dominates $\phi_d$ at late times for a sufficiently long period of sufficiently strong differential rotation.
As discussed above, such a tightening of the spiral is accompanied by a diminishing pressure contrast, while the density and temperature contrasts can remain constant in the absence of radial flows.
If differential rotation subsides into uniform rotation, the integral becomes $r$-independent, and the discontinuity pattern freezes in the corotating frame.

Observations \citep[see discussion in][]{Keshet2012} and simulations \citep[\eg][]{AscasibarMarkevitch06} typically indicate that spirals are trailing, rather than leading.
In the present notations, where we assumed $\partial_r \phi_{d}<0$, this corresponds to $\omega>0$.
Equation \eqref{eq:CFPattern} shows that a trailing spiral is a natural outcome of differential rotation, in which $|\omega(r)|$ typically monotonically decreases.
Indeed, if such a declining $|\omega(r)|$ profile is sustained for a sufficient duration, as expected at large radii, the integral reproduces $\partial_r\phi_d<0$ only if $\omega>0$, \ie when the spiral is trailing.
Note that at very small radii, there may be a central region in which $|\omega(r)|$ increases, rather than decreases, radially.
If such a rising $\omega(r)>0$ profile is sustained for a period long enough to drive $\partial_r\phi_d>0$, a leading spiral can be produced at small scales. Indeed, for some initial conditions, we numerically produce composite spiral structures that are leading at small radii and trailing at large radii.
One may also consider more complicated flows, with $\omega$ changing sign, but henceforth we assume for simplicity that $\omega>0$.

\subsubsection{Thermal spiral also in pressure}
\label{subsec:PSpiral}

As shown in \S\ref{subsec:SelfSimilar}, a spiral structure must manifest also in the pressure distribution; this conclusion is quite general and does not require self-similarity.
Indeed, such a pressure spiral is necessary in order to entrain the mixed ICM, preventing the low (high) density phases from buoyantly rising (sinking).
To see this, consider the leading order \eqref{eq:NSr} of momentum conservation, before incorporating below more subtle effects such as differential rotation and viscosity.
As $\bm{g}$ is radial, we see that $\rho^{-1}\partial_r P$ must be approximately radial, too.
Hence, if $\rho$ shows a spiral structure and thus varies with $\phi$, pressure and its radial gradient $\partial_r P$ must vary with $\phi$, too.
The implied, subtle pressure spiral can be obtained by integrating Eq.~\eqref{eq:NSr} inwards along radial rays,
\begin{equation}\label{eq:PSol}
P(\bm{r}) \simeq \int_\infty^{r} \rho(r',\theta,\phi)g(r')\,dr' \fin
\end{equation}

The inferred properties of the pressure spiral are qualitatively similar to those derived above for the self-similar case.
The pressure field \eqref{eq:PSol} is continuous even at the discontinuity, although its gradient there is not.
As $g$ and $dr$ are negative, each contribution to the inwards integral is positive.
As $\rho$ drops outside the discontinuity, this positive contribution is small (large) just outside (inside) the discontinuity.
We therefore expect that at any constant $r$, the azimuthal pressure profile $P(\phi)$ is minimal near the CF and maximal near the opposite side of the cluster; as shown below, this $P(\phi)$ profile is approximately given by a concave parabola.
As $\partial_\phi P \propto \int g \partial_{\phi}\rho\, dr$, the azimuthal pressure gradient scales with $\delta\rho\propto (q-1)/(q+1)$ and with the distance $\Delta r$ between spiral windings.
Thus, the pressure spiral becomes more subtle as the spiral tightens or the contrast $q$ diminishes.

\begin{figure}
\begin{center}
\myfig{
\includegraphics[width=1.0\linewidth,trim={0cm 0cm 0cm 0cm},clip]{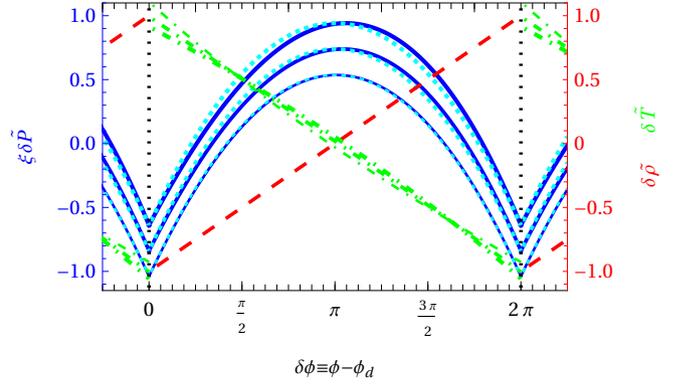}
}
\end{center}
\caption{
Normalised azimuthal profiles $\delta \tilde{A}$ (notations as in Fig.~\ref{fig:SSParabola}) computed from Eq.~\eqref{eq:PSol}.
In this example, we assume a hyperbolic, $\phi_d=20(a/r)$ spiral with the linear $\delta\rho$ profile \eqref{eq:LinearDRho} and a constant contrast $q=1.5$ superimposed on an ICM with a Hernquist profile of scale length $a$.
The azimuthal profiles are shown at radii $r/a=0.5$, $1.0$, and $1.3$ (thin to thick curves; $\delta\tilde{P}$ manually offset upwards for visibility).
The pressure profile is well-fit by a parabola (fitted dotted cyan curves), especially in the inner parts of the spiral, and is multiplied for visibility by the local tightness parameter $\xi$ estimated from Eq.~\eqref{eq:NSrr}.
\label{fig:PParabola}
}
\end{figure}

Given a model for the discontinuity pattern $\phi_{d}(r,\theta)$, for the density distribution $\rho(\bm{r})$, and for $g(r)$, one may directly compute the pressure field \eqref{eq:PSol}, as demonstrated in Fig.~\ref{fig:PParabola}.
Recall that in our notations, the spiral pattern opens clockwise, \ie $\partial_r \phi_d<0$, so (except within the thin discontinuity transition) $\partial_\phi \rho>0$ and $\partial_\phi T<0$.
Furthermore, $\delta\rho\equiv \rho/\rho_0-1$ is approximately linear in $\phi$, as given in Eq.~\eqref{eq:LinearDRho}.
Here, we do not invoke self-similarity, so $\rho_0$ is defined as the azimuthal mean of $\rho$, and $q$ may vary with both $r$ and $\theta$.
As the $P(\phi)$ variations are typically small, a linear $\delta\rho(\phi)$ implies that $T(\phi)$ is also approximately linear (dot-dashed green curves in the figure).

Under these assumptions, $\partial_{\phi\phi}P$ is negative and does not vary much in the co-rotating frame, so $P(\phi)$ is approximately a concave parabola.
Indeed, for the linear $\delta\rho(\phi)$ of Eq.~\eqref{eq:LinearDRho}, twice taking the azimuthal derivative of Eq.~\eqref{eq:PSol} indicates that $\partial_{\phi\phi}P$ is constant along $r$ between discontinuity windings (where $\partial_{\phi\phi}\rho=0$), and only picks up a contribution each time a discontinuity is crossed inwards along the integration path, given by
\begin{align} \label{eq:Pphiphi}
& \Delta(\partial_{\phi\phi}P)_d  =
\partial_{\phi}\Delta(\partial_{\phi}P)_d=
\partial_{\phi}\left(\int_{r_{d+}}^{r_{d-}} g \partial_{\phi}\rho\,dr\right) \\
& \quad\quad \quad = \partial_{\phi}\left[ \int_{\phi_{d+}}^{\phi_{d-}} g{\partial_{\phi}\rho} \, \frac{d(\delta\phi)}{-\partial_r\phi_d} \right]
= \partial_{\phi}\left( \frac{-g}{\partial_r\phi_d}2\rho_0\frac{q-1}{q+1}\right)_d \coma \nonumber
\end{align}
where subscript $d$ designates evaluation at the discontinuity.
Here, we used the limit where the infinitely-thin discontinuity can be equally crossed either radially or azimuthally, with subscript $d+$ ($d-$) refering to just outside (inside) the transition.
In a tight spiral, $\Delta(\partial_{\phi\phi}P)_d$ is negative and slowly varying, because the last brackets are negative and become slightly more negative as $\phi$ increases, \ie as the discontinuity radius slightly decreases.
This conclusion remains intact even in the presence of small corrections to the linear $\rho(\phi)$, and even if $q$ somehow increases radially slowly.

Figure \ref{fig:PParabola} illustrates the thermal spiral obtained from Eq.~\eqref{eq:PSol}.
Here, we assume a hyperbolic spiral pattern with a constant density contrast $q$ in an ICM with a Hernquist distribution.
The figure shows the normalised azimuthal deviations $\delta\tilde{A}$ of each thermal quantity $A$, defined as in Eq.~\eqref{eq:DeltaTildeA}, but here
\begin{equation} \label{eq:deltaADef}
\delta A \equiv \frac{A}{\bar{A}}-1
\end{equation}
is normalised more generally by its azimuthal average,
\begin{equation} \label{eq:AzAvgA}
\bar{A}\equiv(2\pi)^{-1}\int A\,d\phi \fin
\end{equation}
As shown in \S\ref{subsec:SelfSimilar}, in the self-similar regime $\bar{A}=C_A A_0$, so the present definitions coincide with their self-similar counterparts for $\delta\rho$, $\delta P$, and (with the $C_T$ correction of Fig.~\ref{fig:SSTandK}) $\delta T$.
As the figure shows, $P(\phi)$ is approximately a concave parabola, especially at small radii deeper inside the spiral.
For such simple distributions and spiral patterns, the integral can be carried out analytically, after incorporating the spiral discontinuity; see Appendix \S\ref{sec:AppPSpiral}.

\subsubsection{Planar evolution}
\label{subsec:PlanarEv}

Consider the regime where $\bm{v}_1$ can be neglected in Eq.~\eqref{eq:RotationV}, so we may study the flow in 2D, or equivalently in 3D but restricted to a plane parallel to the spiral plane.
Let us assume that in a frame locally co-rotating with the spiral structure at the velocity $\bm{v}=\omega\varrho\unit{\phi}$, the flow evolves slowly and so can be approximated as a steady-state.
If radial flows were present, it would be convenient to analyse the dynamics parallel to the spiral tangential discontinuity surface \citep{Keshet2012}, but in the current case it is advantageous to reduce momentum conservation \eqref{eq:NSFull} into a cylindrically-radial equation,
\begin{equation}\label{eq:NS_r}
\frac{1}{\rho}\frac{\partial P}{\partial \varrho} = f_\varrho
\end{equation}
and an azimuthal equation,
\begin{equation}\label{eq:NS_phi}
\frac{1}{\rho \varrho}\frac{\partial P}{\partial \phi} = f_\phi \fin
\end{equation}
Here, we defined the corresponding specific effective forces,
\begin{equation} \label{eq:f_r}
f_\varrho = g+\omega^2 \varrho + \frac{\partial_\phi \mu}{\rho} \partial_\varrho \omega
\end{equation}
in the $\varrho$ direction, and
\begin{equation} \label{eq:f_phi}
f_\phi = \frac{\partial_\varrho\left(\varrho^3\mu \partial_\varrho \omega\right)}{\varrho^2\rho} - \dot{\omega}\varrho
\end{equation}
in the $\phi$ direction, where we used the shorthand $\dot{\omega}\equiv\partial_t\omega$.

As pressure is continuous across the discontinuity, multiplying Eq.~\eqref{eq:NS_phi} by $\varrho^2\rho$ and integrating over $\phi$ yields
\begin{equation} \label{eq:SpiralEv1}
0 = \partial_\varrho\left(\varrho^3 \partial_\varrho \omega\right)\bar{\mu} + \varrho^3 \partial_\varrho \omega \,\overline{\partial_\varrho \mu} - \dot{\omega}\varrho^3\bar{\rho}  \fin
\end{equation}
In general, $\overline{\partial_\varrho \mu}$ differs from $\partial_\varrho \overline{\mu}$ and is sensitive to the $\mu(T)$ dependence and the structure of the temperature spiral.
However, we recover nearly indistinguishable spiral structures numerically when the temperature dependence of $\mu(T)$ is weakened or even eliminated, whereby we may approximate Eq.~\eqref{eq:SpiralEv1} as
\begin{equation} \label{eq:SpiralEv2}
\partial_\varrho\left(\varrho^3 \bar{\mu}\partial_\varrho \omega\right) \simeq \varrho^3\bar{\rho} \, \dot{\omega} \fin
\end{equation}
This partial differential equation (PDE) approximates the evolution of $\omega$, and hence of the spiral discontinuity, as a function of the underlying $\bar{\rho}(\varrho)$ profile, the $\mu(T)$ profile, and the initial perturbation.
Indeed, given a solution $\omega(t;\varrho)$ to Eq.~\eqref{eq:SpiralEv2}, the integral in the advection Eq.~\eqref{eq:CFPattern} can be carried out, giving an approximate discontinuity pattern $\phi_d(\varrho)$.

One can qualitatively characterize the evolution of $\omega(t;\varrho)$ in relation to the two stationary solutions of Eq.~\eqref{eq:SpiralEv2}: uniform, $\omega_1\propto \varrho^0$ rotation and differential, $\omega_2\propto (\varrho^{2} \bar{\mu})^{-1}$ rotation.
Differential rotation with $\partial_\varrho \omega<0$, as expected at large radii, slows down if it lies between these two solutions, \ie if $\omega(\varrho)$ diminishes as $\varrho$ increases, but slower than $(\varrho^2 \bar{\mu})^{-1}$.
In relaxed regions where the radial dependencies of $\omega$ and $\rho$ are close to power-laws, this temporal decline in $\omega$ is approximately exponential.
If $\partial_\varrho(\rho\varrho/\bar{\mu})<0$, as expected at large radii where $\rho(\varrho)$ declines steeply, the fractional slowdown of $\omega$ is faster as the radius increases, and the solution slows down towards the differential rotation solution $\omega_2$.
For example, for an isothermal sphere, $\rho\propto \varrho^{-2}$ distribution, where $\bar{\mu}$ is constant, an initial $\omega\propto r^{\lambda_\omega}$ profile slows down towards $\omega_2\propto r^{-2}$ for any $-2<\lambda_\omega<0$.
If, in contrast, $\partial_\varrho(\rho\varrho/\bar{\mu})>0$, as may occur in a flat density core or when $\bar{\mu}(T(\varrho))$ declines rapidly, fractional slowdown is faster at smaller radii, and the solution tends towards uniform rotation $\omega_1$.

In practice, the PDE solution is sensitive to the precise initial conditions, and is generally not a power law.
Some specific solutions can be analytically found and integrated to yield the discontinuity pattern.
For example, for an isothermal sphere perturbed at small radii, integrating the PDE solution yields a Lituus spiral, $\phi_{d}\propto \varrho^{-2}$.
More generally, for $\bar{\mu}\propto \varrho^m$ and $\bar{\rho}\propto \varrho^{-2+m}$ with a free parameter $m$, the analytic solution $\omega\propto \tau^{-1/2}e^{-\tau/4}\varrho^{-1-m/2-\ln(\varrho)/(4\tau)}$ to the PDE yields a $\phi_{d}\propto \varrho^{-2-m}$ spiral at late times.
Here, $\tau$ is a rescaled time, and the integral in Eq.~\eqref{eq:CFPattern} was carried out from $t=0$ to $t\to\infty$.
Solving the PDE numerically for a Hernquist density profile perturbed at small radii gives approximately $\phi_{d}\propto \varrho^{-1}$ for $\bar{\mu}\propto \varrho^{-1}$, and $\phi_{d}\propto \varrho^{-2}$ for $\bar{\mu}\propto \varrho^0$.

\begin{figure}
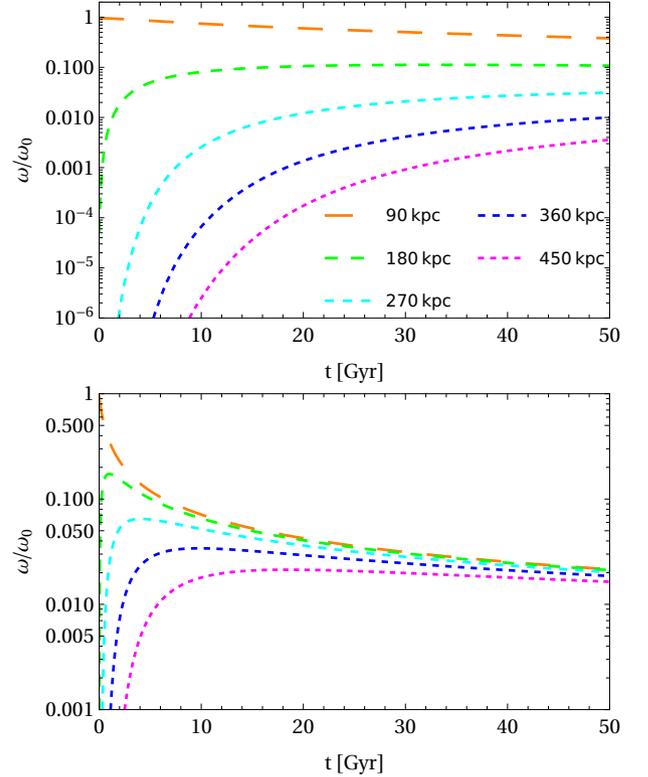

\begin{center}
\myfig{
\includegraphics[width=8cm]{\figeps{OmegaTp0}}
\includegraphics[width=8cm]{\figeps{OmegaTp2.5}}
}
\end{center}
\caption{
Angular frequency evolution $\omega(t)$ at different radii (see legend) according to Eq.~\eqref{eq:SpiralEv2}, starting with initial uniform rotation with $\omega=\omega_0$ inside $r<0.2a$, for our nominal Hernquist profile (total mass $M=10^{15}M_\odot$ and scale length $a=676\kpc$). Spitzer viscosity $\bar{\mu}=\mu_s$ is assumed with a constant temperature $k_B T=3\keV$ (top panel) and with $k_B T =3(a/r)\keV$ (bottom).
}
\label{fig:TheoryOmega}
\end{figure}

The solution $\omega(t; \varrho)$ for a Hernquist density profile with Spitzer viscosity $\bar{\mu}=\mu_s(T)$ is shown in Fig.~\ref{fig:TheoryOmega} for two different temperature profiles, demonstrating the sensitivity of the $\omega$ evolution to viscosity. Here, we adopt the parameters of our nominal simulations set up in \S\ref{subsec:SimSetup}: a total mass $M=10^{15}M_\odot$, a scale length $a=676\kpc$, and a perturbation peaked at $\varrho\simeq 0.1a$; for simplicity, the initial conditions are taken as uniform rotation in the centre, with a strong exponential decay outside $\varrho=0.2a$. As anticipated, in the $\varrho\lesssim a$ core, the solution quickly asymptotes to uniform rotation $\omega_1$ only when the $\bar{\mu}(\varrho)$ profile is declining; differential rotation steeper than $\omega_2$ is found at large radii. Note that $t\propto \mu^{-1}$ in Eq.~\eqref{eq:SpiralEv2}, so globally raising the viscosity simply expedites the $\omega$ evolution by rescaling time.

The pressure derivative $\partial_\phi P$ in the azimuthal momentum equation \eqref{eq:NS_phi} is small, so to low order the equation reduces to Eq.~\eqref{eq:SpiralEv2}.
To higher orders, without averaging Eq.~\eqref{eq:NS_phi} over $\phi$, this equation is in general inconsistent with a concave $P(\phi)$ parabola.
For instance, typically $\dot{\omega}<0$, so the last term in Eq.~\eqref{eq:f_phi} is positive, and as $\rho$ drops across the CF, this term contributes a CF drop in $\partial_\phi P$, rather than the jump anticipated in a concave parabola.
The viscous term in Eq.~\eqref{eq:f_phi} does not remedy the problem, as it is typically positive, and its contribution to $\partial_\phi P$ is independent of $\phi$ if $\mu$ is assumed constant.
Furthermore, inasmuch as the viscous term applies within the discontinuity transition, the jump in $T$ induces a negative spike in $\partial_\phi P$ if the positive temperature dependence of $\mu(T)$ is retained, which is inconsistent with a continuous $P(\phi)$.

Such arguments suggest that at high orders, the small correction term $\bm{v}_1(t;\bm{r})$ in Eq.~\eqref{eq:RotationV} becomes important.
Numerically, we find that time-dependent peculiar velocities, which are not resonant with the spiral structure, persist throughout the evolution.
Indeed, one can show analytically that there is no exact solution to the adiabatic fluid equations that admits a quasi-steady state featuring a purely (differentially) rotating, viable spiral structure, with or without viscosity.
Therefore, while Eq.~\eqref{eq:SpiralEv2} is likely to govern the bulk evolution of rotation and the resulting tightness of the spiral, it is only approximate; additional effects can modify the $\omega$ and $\phi_d$ profiles, and may well drive the spiral towards its self-similar solution.
This possibility is supported by our numerical results, in particular by the similar spiral structures obtained when assuming different $\mu(T)$ dependencies.

\subsubsection{Radial flows and modified convective instability}
\label{subsec:RadFlowAndInst}

While the thermal distribution is governed mainly by the radial hydrostatic equilibrium Eq.~\eqref{eq:NSr}, and thus admits the simple self-similarity scaling \eqref{eq:SSSAnsatz}, the flow pattern does not need to precisely adhere to the same similarity scaling, and in fact must include at least the aforementioned second-order deviation $\bm{v}_1(t;\bm{r})$.
An outflow, an inflow, and a combination of both inflow and outflow separated by the tangential discontinuity, are possible, but are sensitive to effects neglected in Eq.~\eqref{eq:NSr}.
For a discussion of flows along the spiral discontinuity, also taking into account effects such as radiative cooling and feedback, see \citet{Keshet2012, ReissKeshet2015}.

While our self-similar spiral ansatz does not modify the radial, \ie azimuthally-averaged, thermal profiles of density and pressure (see \S\ref{subseubsec:UnchangedRhoP}), it does rescale the other radial profiles, as shown above (see \S\ref{subseubsec:ModifiedTandK}).
Moreover, all thermal profiles can change due to radial flows.
In particular, the deposition of high entropy plasma near the centre, by the initial perturbation or subsequent flows, could render the centre of the cluster convectively unstable.
Such an unstable region, emerging where the \citet{Schwarzschild1958book} instability criterion $\partial_r K<0$ holds, would then mix the plasma and lead to a core of constant entropy, $\partial_r K=0$.

In addition, radial flows or mere changes to $\xi$ or $q$ can induce a local convective instability, further altering the thermal profiles.
Consider such an instability, emerging for example as a tightening spiral with substantial CF entropy jumps renders the radial entropy gradient between discontinuity windings negative, as discussed in \S\ref{subseubsec:ModifiedTandK}.
Even a positive but shallow entropy profile could lead to a local convective instability in a spiral structure, where small motions in the azimuthal direction entail a change in ambient entropy and may thus induce circulation.
In either case, a local $\partial_r K<0$ would lead to a local convective instability within a radial segment confined by the discontinuity both from above and from below, saturating when $K$ becomes constant within such a segment.

Interestingly, a global convective instability may emerge in such a spiral structure, even if $\partial_r K\geq 0$ everywhere.
Consider some region of a spiral structure in which each radial segment develops a constant entropy, $K(r,\phi)=K(\phi)$, over a short timescale due to a local convective instability.
Let us define the average of quantity $A$ along a radial segment of length $l$, confined by the discontinuity both above and below, as
\begin{equation}
\langle A \rangle \equiv \frac{1}{l} \int A \,dl \fin
\end{equation}
Say that such a segment moves outward along the spiral, from $r$ to $r+dr$, where $r$ refers to, say, the midpoint of the segment.
The segment is then stretched to length
\begin{equation} \label{eq:SegmentStretch}
l+dl\simeq l(1+dr/r)\fin
\end{equation}
This approximation agrees with simulations and is precise for a logarithmic spiral, itself a good approximation for observed ICM spirals.
In particular, Eq.~\eqref{eq:SegmentStretch} holds in a hyperbolic spiral up to $O(N+1)^{-1}$ corrections, where $N$ is the number of windings counted inwards from infinity.

The mean pressure in the segment changes, as it rises along the spiral, at a rate
\begin{align} \label{eq:dPSegment}
\left( \frac{d\langle P\rangle}{dr} \right)_s &
\simeq \left( \frac{\partial \langle P\rangle}{\partial \phi} \right)_{s,l} \left( \frac{d\phi}{dr} \right)_{s} + \left( \frac{\partial\langle P\rangle}{\partial l} \right)_{s,\phi} \left( \frac{dl}{dr} \right)_s  \\
& = \frac{\int \partial_\phi P\, dl}{rl} - \frac{ \langle P\rangle - \Delta P_d }{l}\left( \frac{dl}{dr} \right)_s
\simeq \frac{\partial \langle P\rangle }{\partial r} - \frac{\langle P\rangle }{r} \coma \nonumber
\end{align}
where a subscript $s$ denotes a derivative along the spiral, subscripts $r,l,K,\ldots$ denote respectively a fixed radius, segment length, entropy, \ldots, and we used Eq.~\eqref{eq:SegmentStretch} to approximate $(dl/dr)_s\simeq l/r$.
The term $\Delta P_d$, accounting for the pressure at the endpoints of the segment, where it touches the discontinuity, was dropped in the last step assuming one of several circumstances: a segment that moves while keeping its endpoints fixed, temporary force imbalance at the endpoints as they move, strong non-thermal pressure, \eg in a magnetic layer near the discontinuity, or a strongly concave $P(\phi)$ being minimal at the discontinuity.

As the pressure declines in the rising segment, adiabatic expansion lowers its density.
An instability would ensue if the segment becomes buoyant, \ie if the ambient density in its new location is higher than its new density,
\begin{align}
\frac{\partial \langle \rho \rangle}{\partial r} > \left(\frac{d\langle \rho \rangle}{d\langle P \rangle}\right)_{K} \left( \frac{d\langle P\rangle }{dr} \right)_s
& \simeq  \frac{\langle \rho\rangle }{\Gamma \langle P\rangle }\left( \frac{\partial \langle P\rangle }{\partial r} - \frac{\langle P\rangle }{r}\right)  \nonumber \\
& \simeq \frac{\langle \rho \rangle}{\Gamma r}\left(\lambda_K-1\right) + \frac{\partial \langle \rho \rangle}{\partial r} \fin
\end{align}
This condition yields a modified Schwarzschild criterion,
\begin{equation} \label{eq:GlobalConvectiveInstability}
\lambda_K\equiv \frac{\partial \ln \bar{K}}{\partial \ln r} \simeq \frac{\partial \ln \langle K \rangle}{\partial \ln r} < 1 \coma
\end{equation}
for a global instability in regions where $\bar{K}(r)$ is sub-linear.
It is tacitly assumed that segments can move around each other (in 3D) or cross each other (even in 2D).

We conclude that a spiral structure that reaches $\partial_r K<0$ locally and $\lambda_K<1$ globally would lead to a local convective instability, saturating with $\partial_r K=0$ within radial segments on a short timescale, followed by a global instability on a long timescale.
At sufficiently small radii, a $\lambda_K=0$ constant entropy core may develop.
At larger radii with an initial $\lambda_K<1$, the global instability would retain $\partial_r K=0$ locally within radial segments, but lead to oscillations in a non-monotonic $\bar{K}(r)$, inducing radial flows that would self-stabilize once $\bar{K}\propto r$ is established.
After the spiral structure dissipates, the $K(r,\phi)=K(\phi)$ segments would disperse, but a local $\lambda_K=1$ would remain imprinted on regions which once harboured a $\lambda_K<1$ spiral.
This behaviour, based on the above effective picture of radial segments moving along the spiral with negligible $\Delta P_d$, appears to be consistent with simulations, as shown in \S\ref{sec:Simulations}.

\section{Numerical simulations}
\label{sec:Simulations}

Merger simulations have long demonstrated the emergence of spiral structure in the ICM \citep{TittleyHenriksen05,AscasibarMarkevitch06}.
We examine a wide range of simulations, of both offset (between baryons and dark matter, as defined below) and merger clusters, in both 2D and 3D, using both Lagrangian and Eulerian codes, with various initial conditions and different forms of viscosity.
For simplicity, baryons are modelled as an ideal fluid with viscosity and with an ideal gas equation of state, without incorporating radiative cooling, AGN feedback, magnetic fields, and transport processes such as particle diffusion and heat conduction.
Consequently, our simulated spirals --- like all other spirals simulated todate --- fail to reproduce the fast outflows inferred from observations along CFs, which may play an important role in speeding up the spiral evolution and regulating its properties.
For this reason, and in order to elucidate the asymptotic, quasi-steady solution, we evolve the ICM over very long durations, exceeding the Hubble time; nevertheless, while noisy, simulated $\sim$ few Gyr spirals already agree qualitatively with our model and show indications for self-similarity.

\subsection{Set-up}
\label{subsec:SimSetup}

For the unperturbed state of the galaxy cluster at the initial time $t=0$, we adopt a spherical Hernquist profile for both dark matter (subscript $dm$) and baryons (subscript $b$, usually omitted), utilizing its rapidly converging mass at large radii.
The mass density $\rho_j$ of each component $j$ is then related to its integrated, conserved mass $M_j$ by
\begin{equation}\label{eq:HQrho}
\rho_j(r)= \frac{M_j a}{2\pi r(r+a)^3} \fin
\end{equation}
All components are assumed to have the same scale length $a$, so the total (subscript $t$) mass density initially satisfies  $\rho_t=\rho_{dm}+\rho_b=\rho_b/f_b=\rho_{dm}/(1-f_b)$ locally.
The gravitational potential then becomes
\begin{equation}\label{eq:HQpotential}
\Phi(r)=-\frac{M_{t}G}{(a+r)}\coma
\end{equation}
where $M_{t}=M_{dm}+M_b$ is the total mass,
implying the hydrostatic-equilibrium baryon temperature profile
\begin{align}
    k_B T  =
    & \bar{m}\sigma^2  
    =
    \frac{GM_{t}\bar{m}}{12a}\left\{\frac{12r(r+a)^3}{a^4}\text{ln}\left(\frac{r+a}{r}\right)\right.\\
    & \quad \quad \quad \left.-\frac{r}{r+a}\left[25+52\frac {r}{a} +42 \left(\frac {r}{a}\right)^2 +12 \left(\frac {r}{a}\right)^3\right]\right\}\nonumber
    \coma
\end{align}
where $\sigma$ is the thermal velocity dispersion (rms velocity in a given direction)
and $G$ is Newton's gravitational constant.
We adopt $M_t=10^{15}M_\odot$ and $a=676\kpc$ as our nominal cluster parameters.

We consider two different types of perturbations: offsetting baryons from the dark matter in position and in velocity, and gravitational off-axis mergers.
Both types of perturbations are set up with angular momentum in the $+\unit{z}$ direction.
A wide range of cluster and perturbation parameters is explored for each scenario.

In the first method, baryons within a radius $r_{\Delta}$ from the centre of the cluster are uniformly displaced spatially by a distance $\Delta x$ in the $+\unit{x}$ direction, and given an initial velocity $\Delta v$ in a perpendicular, $+\unit{y}$ direction.
We adopt $\Delta x=+0.1a$, $\Delta v=+450\km\se^{-1}$, and $r_{\Delta}=2\Mpc$ as our nominal offset parameters, with an exponential drop in the initial $\Delta x$ and $\Delta v$ beyond $r_{\Delta}$.

In the second, merger-type perturbation, a dark matter clump of mass $M/f_M$ crosses the ICM, with radius $r_p$ and velocity $v_p$ at pericentre passage.  We adopt $f_M=5$, $r_p=0.5a$, and $v_p=1000\km \se^{-1}$ as our nominal merger parameters, with pericentre passage offset spatially in the $+\unit{x}$ direction with velocity in the $+\unit{y}$ direction.
For simplicity, in these simulations we use a linear trajectory of a baryon-free subhalo, instead of a self-consistent trajectory of a gravitating gas and dark matter clump; among the gravitating simulations in the literature, our nominal set-up is comparable to the main set-up of \citet[][their section 3]{AscasibarMarkevitch06}.

We carry out Lagrangian simulations using the $N$-body/smoothed particle hydrodynamics (SPH) code GADGET2 \citep{SpringelEtAl01, Springel05} in two (henceforth 2D-GADGET) and three (3D-GADGET) dimensions, and Eulerian simulations using the magnetohydrodynamics code Athena++ \citep{StoneEtAl20} in three dimensions (henceforth Athena, for brevity).
The Lagrangian runs capture the spiral structure including its discontinuities very efficiently \citep{AscasibarMarkevitch06}, in spite, and in part owing to, the inaccurate treatment \citep[][and references therein]{Tricco19} of discontinuity dissipation by such SPH codes, which may coincidentally mimic the isolating effect of the magnetic layers or other physical effects that require an inaccessibly high resolution.
The Eulerian runs reach very high resolutions, using static and adaptive mesh refinement, and allow for an easy control of viscosity.
The combination of the two codes is useful both for confirming the robustness of the results and for testing the underlying assumptions.

For example, we use Lagrangian simulations of dark matter and baryons to test if the effects of an evolving dark matter halo can be approximated as a rigid gravitational potential.
Such a rigid potential was previously shown to successfully approximate merger simulations \citep{ZuHoneRoediger12}; in \S\ref{subsec:QuasiSpiral}, we show that a fixed rigid potential provides a similarly accurate approximation for offset simulations.
Both GADGET and Athena codes are slightly modified, in particular to introduce rigid potentials and physical viscosity.

\begin{bfigure*}
\begin{center}
\vspace{0.5cm}
\begin{overpic}[height=4.1cm]{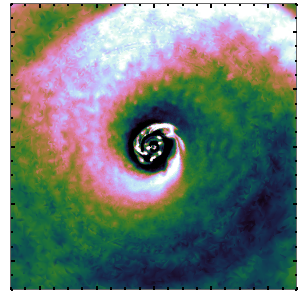}
\put(35,105){$t=1\Gyr$}
\end{overpic}
\begin{overpic}[height=4.1cm]{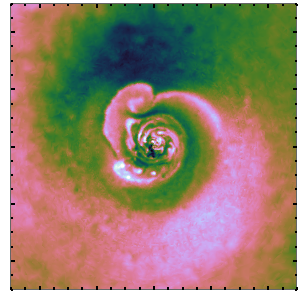}
\put(35,105){$t=3\Gyr$}
\end{overpic}
\begin{overpic}[height=4.1cm]{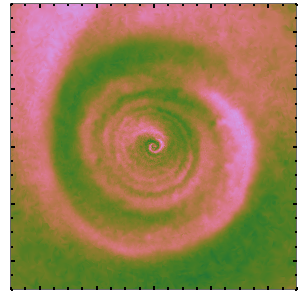}
\put(35,105){$t=10\Gyr$}
\end{overpic}
\begin{overpic}[height=4.1cm]{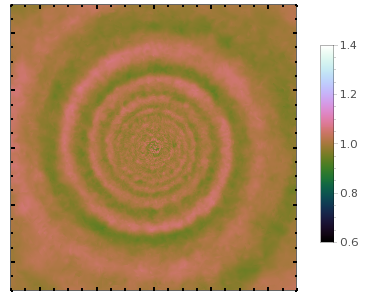}
\put(30,86){$t=30\Gyr$}
\end{overpic}\\
\includegraphics[height=4.1cm]{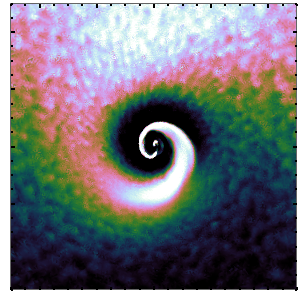}
\includegraphics[height=4.1cm]{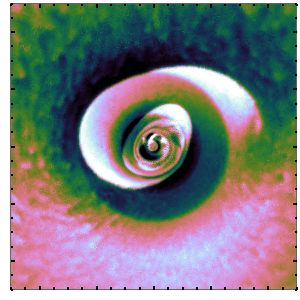}
\includegraphics[height=4.1cm]{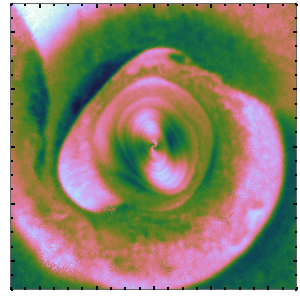}
\includegraphics[height=4.1cm]{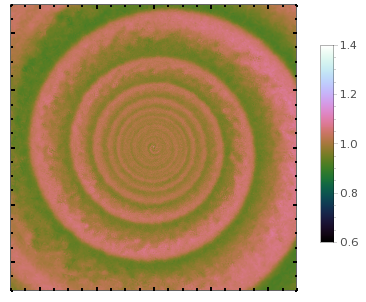}\\
\includegraphics[height=4.1cm]{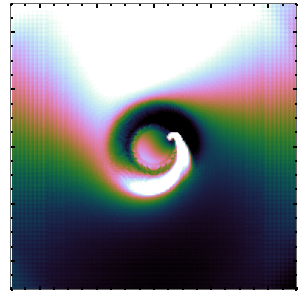}
\includegraphics[height=4.1cm]{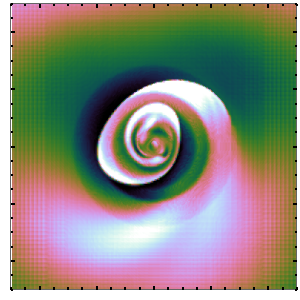}
\includegraphics[height=4.1cm]{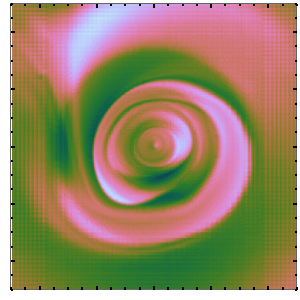}
\includegraphics[height=4.1cm]{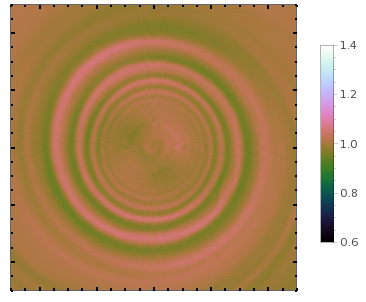}
\end{center}
\caption{
Temporal evolution of nominal offset simulations in 2D-GADGET (top row), 3D-GADGET (middle row), and Athena (bottom).
The density field $\rho/\bar{\rho}$ is shown normalised to its azimuthal mean, in a $1\Mpc$ slice of the spiral plane, at different times (from left to right: $1$, $3$, $10$, and $30$ Gyr). The evolution may be accelerated by processes neglected here, see \S\ref{subsec:QuasiSteadyState}.
}
\label{fig:DifferentEvolution}
\end{bfigure*}

Our nominal Lagrangian simulations are carried out both in 3D and in 2D, using $N_g=10^6$ gas particles in a large cube or square of length $L=200\Mpc$.
The corresponding, nominal gas mass resolution is $f_b M_t/N_g\simeq 1.7\times 10^{8}M_\odot$ in 3D, and $f_b M_t/(2a N_g)\simeq 1.3\times 10^{5}M_\odot\kpc^{-1}$ in 2D.
Runs with particle numbers in the range $10^5\leq N_g\leq 10^8$ are examined, as necessary and in order to demonstrate convergence, giving a maximal mass resolution of $1.7\times 10^{6}M_\odot$ in 3D and $1.3\times 10^{4}M_\odot\kpc^{-1}$ in 2D.
The 3D runs use either gravitating particles, including both gas and $N_{dm}\simeq N_g$ dark matter particles, or only non-gravitating gas particles in a rigid potential.
The 2D runs are limited to non-gravitating gas particles in a periodic simulation square with a rigid potential.
Spiral discontinuities are found to form rapidly and dissipate very slowly even when using only the inherent numerical viscosity and standard artificial viscosity included in the code, with no need to incorporate additional viscosity.

Our Eulerian runs are purely hydrodynamic, simulating gas evolving in 3D under a rigid potential with periodic boundary conditions imposed in each direction.
The nominal runs use static (only) mesh refinement, starting with a level $n=0$ base grid of $128$ cells in each dimension, representing a cube of size $(16\Mpc)^3$.
Logarithmically spaced
refinement levels $n=1,2,3,\ldots$ each halve the grid spacing within a cube of length $L_n=L_{n-1}/2$, reaching level $n=7$ within the $|x,y,z|<3\kpc$ cube for a maximal spatial resolution of $\sim 0.98\kpc$
(for technical reasons, we use $L_5/L_6=4$ instead of 2).
Runs ranging from one-fourth to four times the nominal resolution in each dimension are carried out, for a maximal resolution of $\sim 0.25\kpc$ near the centre.
Nominal runs use \citet{Spitzer} viscosity,
\begin{equation}
\label{eq:Spitzer}
\muS(k_BT) \simeq 5500\left(\frac{T}{10^8 \K}\right)^{5/2}
\gr\se^{-1}\cm^{-1} \coma
\end{equation}
implemented for efficiency only in the central, $r_\mu\simeq 2\Mpc$, falling exponentially to numerical viscosity at larger radii.

\begin{bfigure*}
\begin{center}
\vspace{0.5cm}
\begin{overpic}[height=4.1cm]{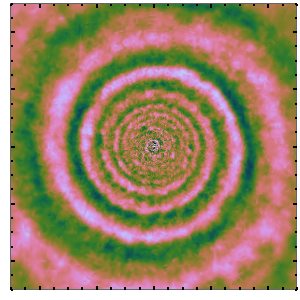}
\put(45,105){$\rho/\bar{\rho}$}
\end{overpic}
\begin{overpic}[height=4.1cm]{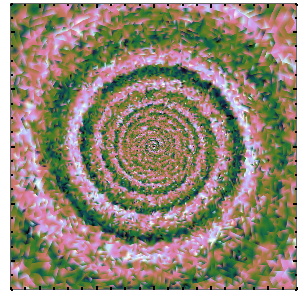}
\put(45,105){$T/\bar{T}$}
\end{overpic}
\begin{overpic}[height=4.1cm]{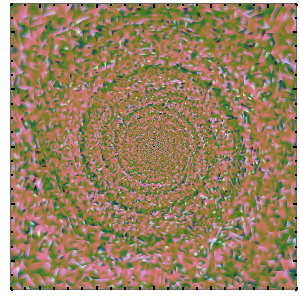}
\put(45,105){$P/\bar{P}$}
\end{overpic}
\begin{overpic}[height=4.1cm]{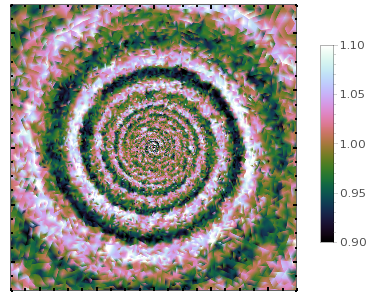}
\put(36,86){$K/\bar{K}$}
\end{overpic}\\
\includegraphics[height=4.1cm]{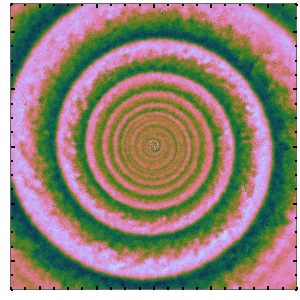}
\includegraphics[height=4.1cm]{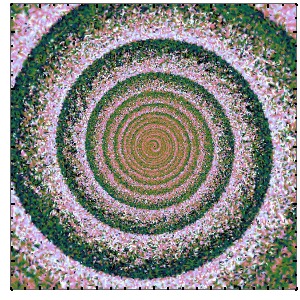}
\includegraphics[height=4.1cm]{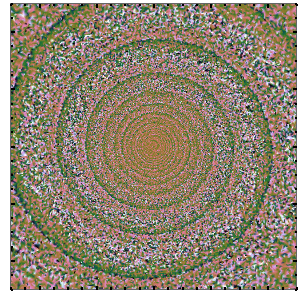}
\includegraphics[height=4.1cm]{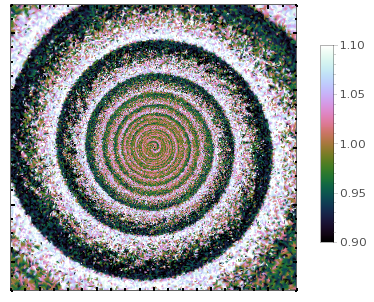}\\
\includegraphics[height=4.1cm]{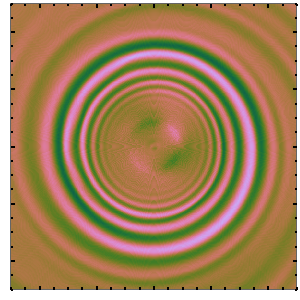}
\includegraphics[height=4.1cm]{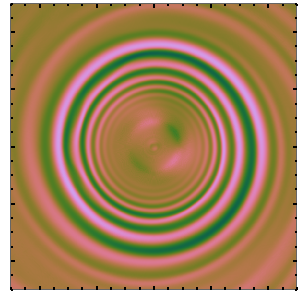}
\includegraphics[height=4.1cm]{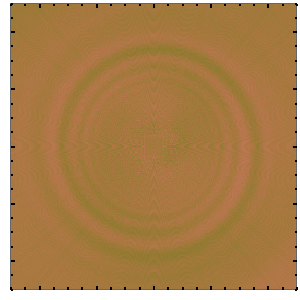}
\includegraphics[height=4.1cm]{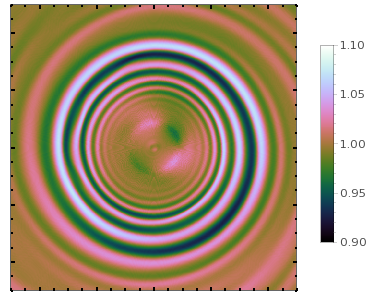}
\end{center}
\caption{
Late-time quasi-steady state in the spiral plane.
Thermal variations $A/\bar{A}$ are shown normalised to their azimuthal mean at $t=40\Gyr$, for offset nominal simulations in a $1\Mpc$ slice.
Columns (left to right): density, temperature, pressure, and adiabat.
Rows (top to bottom): 2D-GADGET, 3D-GADGET, and Athena.
}
\label{fig:SSS_parallel}
\end{bfigure*}

\begin{figure*}
\begin{center}
\vspace{0.5cm}
\begin{overpic}[height=4.1cm]{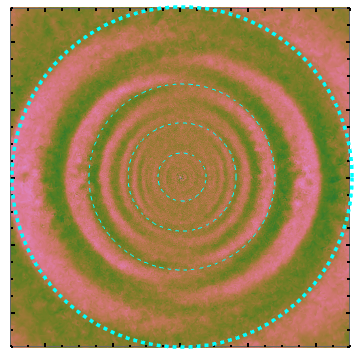}
\put(45,105){$\rho/\bar{\rho}$}
\end{overpic}
\begin{overpic}[height=4.1cm]{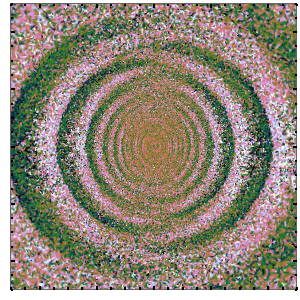}
\put(45,105){$T/\bar{T}$}
\end{overpic}
\begin{overpic}[height=4.1cm]{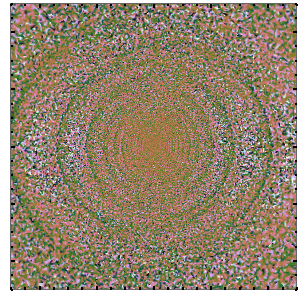}
\put(45,105){$P/\bar{P}$}
\end{overpic}
\begin{overpic}[height=4.1cm]{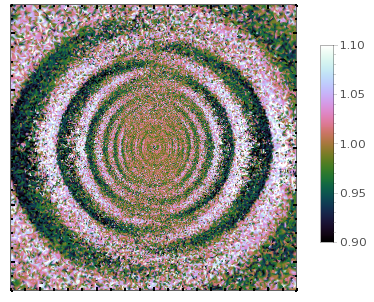}
\put(36,86){$K/\bar{K}$}
\end{overpic}\\
\includegraphics[height=4.1cm]{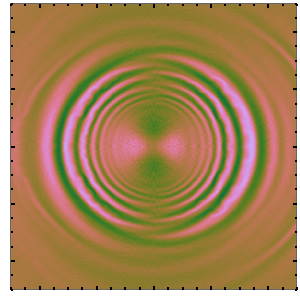}
\includegraphics[height=4.1cm]{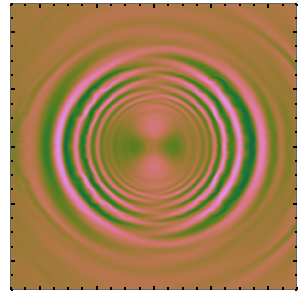}
\includegraphics[height=4.1cm]{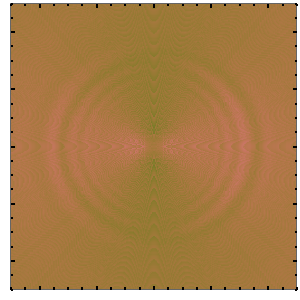}
\includegraphics[height=4.1cm]{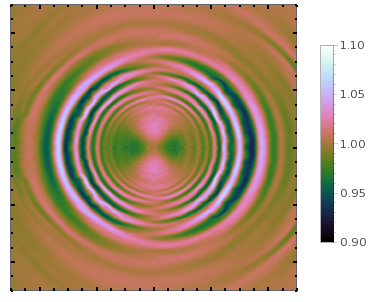}
\end{center}
\caption{
Same as Fig.~\ref{fig:SSS_parallel} but perpendicular to the spiral plane, for 3D-GADGET (top row) and Athena (bottom).
A few concentric circles are plotted in the top-left panel to show that the discontinuities are approximately semicircles.
}
\label{fig:SSS_perp}
\end{figure*}

\subsection{Emergence of a quasi-steady state}
\label{subsec:QuasiSteadyState}

While the early-time evolution of the perturbed ICM depends on the type and parameters of the perturbation, the number of spatial dimensions, and even on the numerical properties of the code, we find that the late-time ICM converges in all cases onto a qualitatively similar spiral quasi-steady state, as demonstrated for the nominal offset simulations in Fig.~\ref{fig:DifferentEvolution}.
As the figure shows, the density evolution in 2D-GADGET, 3D-GADGET and Athena differs substantially at early times, especially on small scales, but by $t\sim 30\Gyr$, the simulations show a fairly similar spiral structure.

Although some differences are still found between different late-time configurations, as seen for example in the right column of Fig.~\ref{fig:DifferentEvolution}, these variations are mostly associated with different levels of spiral tightness, induced by the different properties of viscosity in each simulation.
In these simulations, the quasi-steady state stabilizes around $t\sim 20\Gyr$, with modest subsequent evolution.
This timescale depends on the perturbation and on viscosity, and in more realistic scenarios should become substantially shorter due to cooling, radial flows, and additional physical processes such as magnetic layers isolating the discontinuities.

The late-time structure is broadly consistent with the self-similar model of \S\ref{subsec:SelfSimilar}: a combination of a trailing, hyperbolic spiral in the equatorial plane with semicircles in perpendicular planes, rotating differentially and eventually saturating to uniform rotation, with only small velocities in the co-rotating frame.
This quasi-spiral structure consists of mixed gas phases delineated by a spiral contact discontinuity, with a subtle pressure spiral locking-in the different phases and preventing them from sinking or buoyantly rising.
The azimuthal density and temperature profiles are found to be approximately linear, while the pressure profile is approximately a concave parabola.
The azimuthally-averaged radial profiles show modest changes after the structure has formed, with evidence for both local and global convection instabilities where the spiral was tight.
Overall, these properties are consistent with the modelling in \S\ref{subsec:SelfSimilar} and \S\ref{subsec:NonSelfSimilar}.
In what follows, we describe different aspects of the emerging quasi-steady state.

\subsection{Quasi-spiral structure}
\label{subsec:QuasiSpiral}

The late time, $t=40\Gyr$ distributions of density, temperature, pressure, and entropy in nominal offset simulations are shown below, within the equatorial, $x$--$y$ plane in Fig.~\ref{fig:SSS_parallel}, and within a perpendicular, $x$--$z$ plane in Fig.~\ref{fig:SSS_perp}.
In order to present the full spiral structure within a $1\Mpc$ slice, across which some quantities $A$ vary over several orders of magnitude, such figures show the local $A$ normalised to its average at the same radius within the plane.
Namely, Figs.~\ref{fig:DifferentEvolution} and \ref{fig:SSS_parallel} depict $A(x,y)/\bar{A}(\varrho)$, where $\bar{A}(\varrho)$ is the azimuthal average of $A$ in the $x$--$y$ plane defined in Eq.~\eqref{eq:AzAvgA}.
The perpendicular structure in Fig.~\ref{fig:SSS_perp} is shown using the analogous quantity $A(x,z)/\bar{A}(\varrho_{xz})$, with $\varrho_{xz}$ defined similarly as the radius in the perpendicular, $x$--$z$ plane.

\begin{bfigure*}
\begin{center}
\myfig{
\includegraphics[width=5.6cm]{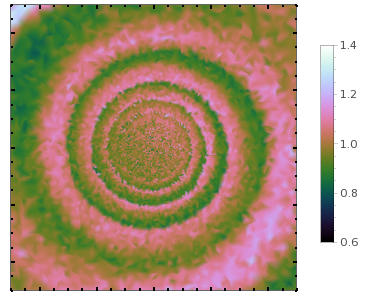}
\includegraphics[width=5.6cm]{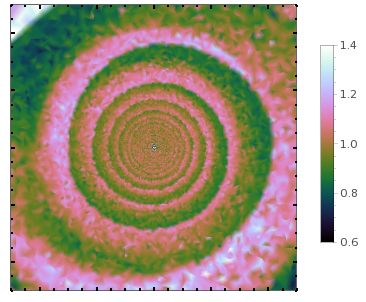}
\includegraphics[width=5.6cm]{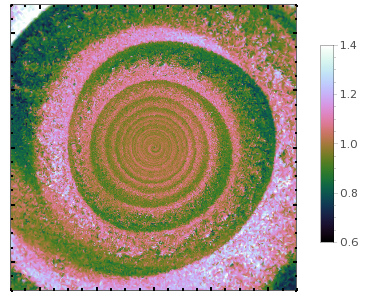}
}
\end{center}
\caption{
Convergence of offset 3D-GADGET simulations: at nominal resolution with DM (left panel) vs. a rigid potential (middle panel), and at high resolution with a rigid potential (right panel). Images show the normalised entropy profile at $t=20\Gyr$ in a $1\Mpc$ box. The spatial offset $\Delta x$ was doubled in the simulation with DM in order to produce roughly the same structure.}
\label{fig:3DOffsetGADGET}
\end{bfigure*}

\begin{bfigure*}
\begin{center}
\myfig{
\includegraphics[width=5.6cm]{\figeps{AthenaK40Low2Res}}
\includegraphics[width=5.6cm]{\figeps{AthenaK40LowRes2}}
\includegraphics[width=5.6cm]{\figeps{AthenaK40HR1}}
}
\end{center}
\caption{
Dependence of offset Athena runs upon resolution: low (left panel), nominal (middle panel), and high (right panel). Images show normalised temperature at $t=40\Gyr$ in a $1.6\Mpc$ slice, for $\mu=3\muS(3\keV)$.
}
\label{fig:AthenaRes}
\end{bfigure*}

The quasi-spiral structure emerges at late times even in low-resolution simulations, although as the resolution is degraded, the central spiral windings are gradually erased as the resolution is degraded and, in Athena, the discontinuity is smoothed out considerably.
As the resolution improves, additional windings appear towards the centre, and the discontinuities become sharper, as seen in the resolution tests for offset simulations of 3D-GADGET in Fig.~\ref{fig:3DOffsetGADGET}, and of Athena in Fig.~\ref{fig:AthenaRes}.
Figure \ref{fig:3DOffsetGADGET} also compares a simulation with dynamical dark matter against a comparable simulation with a rigid potential.
As dark matter motions dissipate some of the structure, we find that doubling the initial offset $\Delta x$ in the simulation with dark matter leads to a late-time structure similar in both simulations.

\begin{bfigure*}
\begin{center}
\myfig{
\vspace{0.5cm}
\begin{overpic}[height=3.5cm]{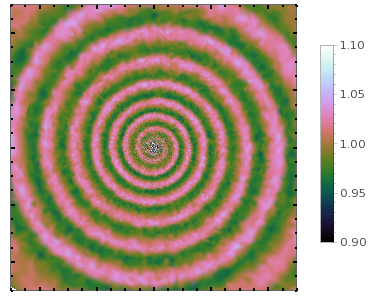}
\put(35,85){$\rho/\bar{\rho}$}
\end{overpic}
\begin{overpic}[height=3.5cm]{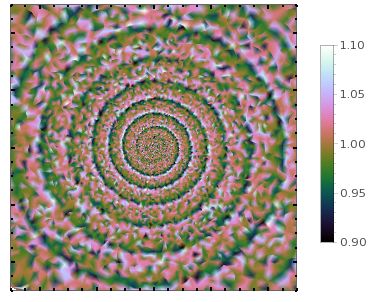}
\put(35,85){$P/\bar{P}$}
\end{overpic}
\begin{overpic}[height=3.5cm]{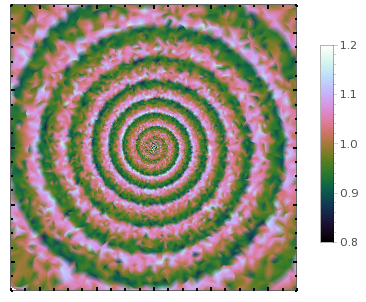}
\put(35,85){$K/\bar{K}$}
\end{overpic}
\begin{overpic}[height=3.5cm]{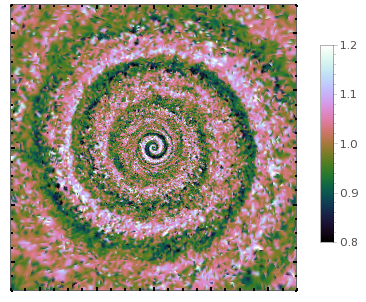}
\put(20,85){$K/\bar{K}$ in 2D}
\end{overpic}\\
\includegraphics[height=3.5cm]{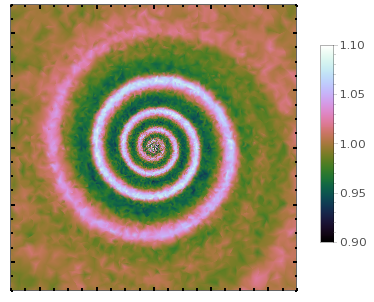}
\includegraphics[height=3.5cm]{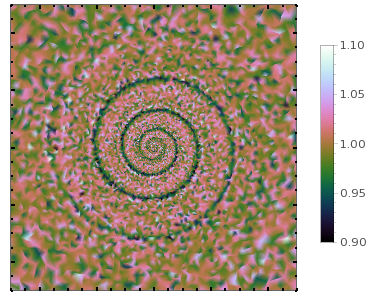}
\includegraphics[height=3.5cm]{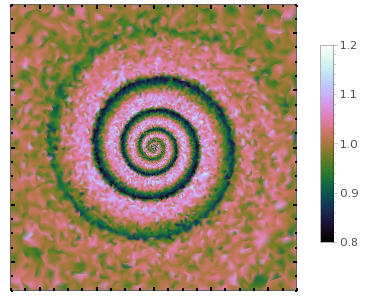}
\includegraphics[height=3.5cm]{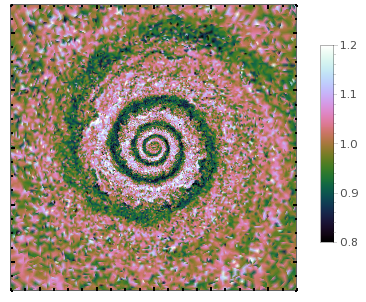}\\
\includegraphics[height=3.5cm]{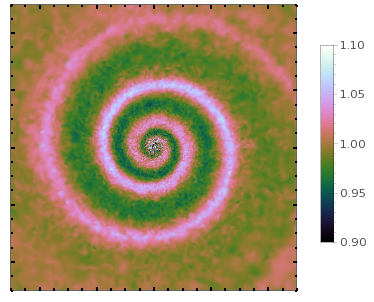}
\includegraphics[height=3.5cm]{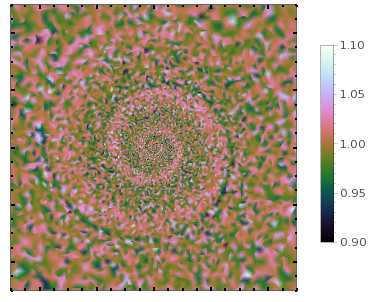}
\includegraphics[height=3.5cm]{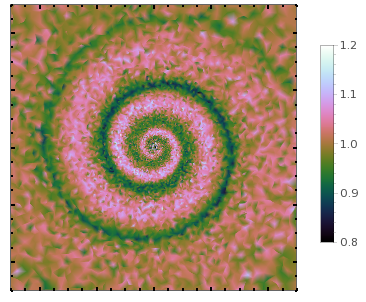}
\includegraphics[height=3.5cm]{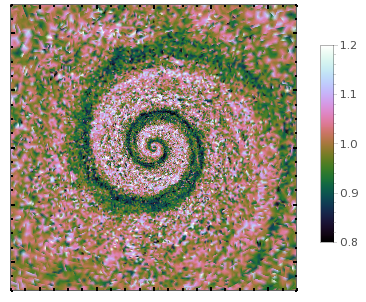}\\
}
\end{center}
\caption{
Late time, $t=40\Gyr$ structure in merger simulations in a $1\Mpc$ slice of the spiral plane.
Columns (left to right): normalised density, pressure, and entropy in 3D-GADGET, and normalised entropy in the corresponding 2D-GADGET.
Rows (top to bottom): nominal, minor ($f_M=10$), and both minor and distant ($f_M=10$ and $b=a$) mergers.
}
\label{fig:MergerRuns}
\end{bfigure*}

Similar late-time spiral structures, albeit with different tightness levels, emerge robustly over a wide range of offset parameters, $\Delta x$, $\Delta v$, and $r_{\Delta}$.
Such spirals form over a somewhat wider parameter range in 3D; the nascent spirals in 2D are more easily disrupted by irregular sloshing for some parameters.
The late-time distribution does not change much as $r_\Delta$ increases beyond $a$, whereas smaller values involve less offset mass and lead to more compact spirals, which never extend to large radii.
For some parameter choices, a small inverted, \ie leading, spiral forms in the centre, within the larger trailing spiral, as anticipated in \S\ref{subsec:AdvectionDiscontinuity}.
Such inverted spirals can persist to late times in 2D, but in general dissipate in 3D and do not survive to late times.

Merger simulations behave similarly to the offset simulations, evolving into a comparable quasi-spiral configuration at late times.
Figure \ref{fig:MergerRuns} demonstrates three such simulations, all showing a rotating, trailing, hyperbolic spiral at late times, including a subtle pressure spiral with minimal $P(\phi)$ along the discontinuity.
As the figure demonstrates, minor mergers or larger impact factors in general yield looser spirals.
In 2D, the spirals are somewhat looser than in 3D, especially in the nominal set-up, in which a small inverted spiral persists in 2D, but dissipates in 3D.

\subsection{Spiral evolution}
\label{subsec:SpiralEvolution}

The late-time configurations found in our various offset and merger simulations differ mainly in the tightness of the spiral and in the degree of ICM differential rotation.
These two diagnostics are related to each other, as discussed in \S\ref{subsec:NonSelfSimilar}, and both are largely controlled by the properties of viscosity in the simulation.
Numerical viscosity is strong and difficult to control in GADGET2, but is sufficiently weak in Athena for us to examine a wide range of physical viscosity properties.
In particular, we replace the temperature dependent $\muS(k_BT)$ by a fixed viscosity, parameterized as $\muS(k_BT_\mu)$ with a constant temperature $T_\mu$, and/or multiply viscosity by a global constant.

\begin{figure*}
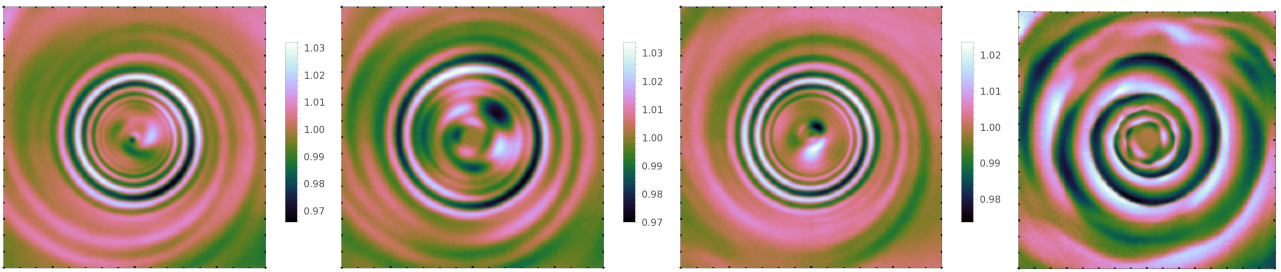

\begin{center}
\myfig{
\includegraphics[width=4.35cm]{\figeps{AthenaK40Sp1Dyn}}
\includegraphics[width=4.35cm]{\figeps{AthenaK40Sp1Fix2}}
\includegraphics[width=4.35cm]{\figeps{AthenaK40LowRes2}}
\includegraphics[width=4.35cm]{\figeps{AthenaK40NoVisc}}
}
\end{center}
\caption{
Dependence of offset Athena runs upon viscosity (left to right): $\mu=\muS$, $\muS(3\keV)$, $3\muS(3\keV)$, or numerical viscosity only. Images show normalised temperature at $t=40\Gyr$ in a $1.6\Mpc$ slice at nominal resolution.
}
\label{fig:AthenaViscosity}
\end{figure*}

Figure \ref{fig:AthenaViscosity} shows the dependence of late-time nominal Athena offset simulations upon viscosity:
the nominal $\mu=\muS(k_BT)$, modified $\muS(3\keV)$ and $3\muS(3\keV)$, and numerical viscosity only (top to bottom panels).
As the last panel shows, numerical viscosity alone is sufficient to generate a spiral, although it is weak, loose, and irregular.
Uniformly strengthening the viscosity leads to a tighter and more regular late-time spiral.
Fixing viscosity uniformly with $T_\nu=3\keV$ renders the spiral less regular and somewhat looser than for nominal viscosity.
The combination of both fixed and strengthened viscosity $\mu=3\muS(3\keV)$ produces results similar to the nominal case.
As uniform viscosity is easier to model analytically (see \S\ref{sec:Model}), this enhanced, uniform prescription is studied below in some detail and referred to as our fixed-viscosity Athena simulations.

\begin{figure}
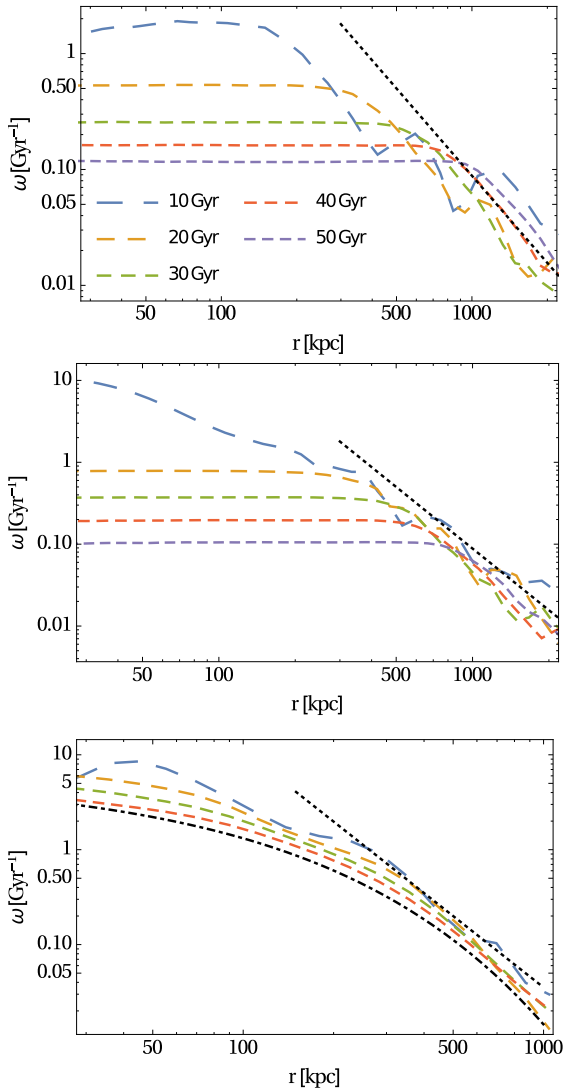

\begin{center}
\myfig{
\includegraphics[width=7.3cm]{\figeps{OmegaOfR2D}}
\vspace{0.2cm}
\includegraphics[width=7.3cm]{\figeps{OmegaOfR3D}}
\vspace{0.2cm}
\includegraphics[width=7.3cm]{\figeps{OmegaOfRAthGal}}
}
\end{center}
\caption{
The radial profile $\omega(r)$ of angular frequency measured in the spiral plane at different times (shorter dashing for later time; see legend in top panel) in 2D GADGET (top panel), 3D GADGET (middle), and Athena (bottom; $\mu=2\muS(3\keV)$) simulations.
At large radii, all simulations approximately show $\omega\propto r^{-5/2}$ (dotted black curve).
At smaller radii, due to their different viscosities, GADGET simulations show uniform rotation, whereas Athena roughly yields $\omega\propto \exp [-(r/50\kpc)^{0.6}]$ (dot-dashed black).
}
\label{fig:omegaVSr}
\end{figure}

\begin{figure}
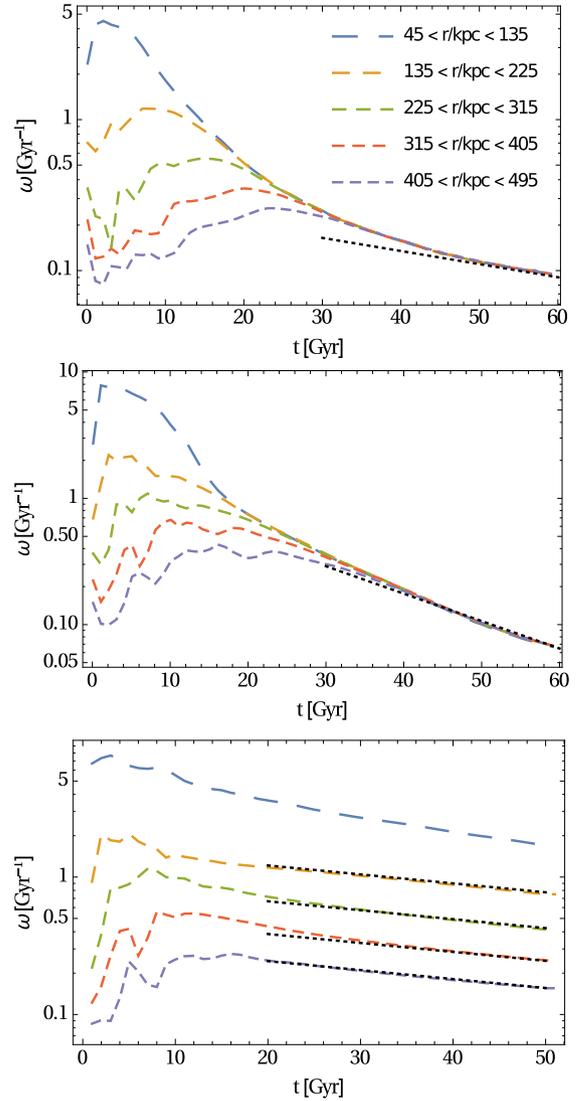

\begin{center}
\myfig{
\includegraphics[width=7.3cm]{\figeps{OmegaOftGad2D}}
\vspace{0.2cm}
\includegraphics[width=7.3cm]{\figeps{OmegaOftGad3D}}
\vspace{0.2cm}
\includegraphics[width=7.3cm]{\figeps{OmegaOftAthMu2}}
}
\end{center}
\caption{
The temporal evolution of $\omega$ measured in the spiral plane at different radial bins (shorter dashing for larger radii; see legend) in 2D GADGET (top panel), 3D GADGET (middle), and Athena (bottom; $\mu=2\muS(3\keV)$) simulations.
At late times, $\omega(t)$ is approximately exponential (dotted black curve).
}
\label{fig:omegaVSt}
\end{figure}

The advection of the discontinuity, as given by Eq.~\eqref{eq:CFPattern}, indicates that the spiral pattern is affected by the rotation pattern and its evolution: the spiral tightens for stronger or longer-lasting differential rotation.
The properties and evolution of differential rotation are, in turn, determined by the viscosity, as shown in \S\ref{subsec:PlanarEv}, and therefore differ among our simulations.
The differential rotation profile in the spiral plane is illustrated for different simulations, as a function of radius at different times in Fig.~\ref{fig:omegaVSr}, and as a function of time for different radial bins in Fig.~\ref{fig:omegaVSt}.

As the figures show, the rotation, initially limited to small radii, gradually spins up increasingly more distant spherical shells.
This coupling to larger radii is mediated to some extent by the initial bulk flows induced by the perturbation, but is mostly driven by viscosity.
This dominant role of viscosity in transferring angular momentum is indicated by the similar spin-up of gas inside and outside of the spiral plane, and by the variation in rotation patterns among the different simulations, which share similar bulk flows but differ in their viscosity properties.

At late times, $\omega(t)$ declines approximately exponentially (dotted lines in Fig.~\ref{fig:omegaVSt}), with differential rotation properties that depend on viscosity, as anticipated in \S\ref{subsec:PlanarEv}.
While Athena sustains differential rotation throughout the simulation, and thus increasingly winds up the spiral, the viscosity in GADGET is sufficiently strong to dissipate the differential rotation and freeze the spiral pattern in a uniformly rotating ICM.
At large radii, beyond the uniform rotation of GADGET, all simulations show differential rotation that approximately follows an $\omega\propto r^{-5/2}$ power law (dotted lines in Fig.~\ref{fig:omegaVSr}), close to the  $\omega_2\propto (r^{2} \bar{\mu})^{-1}$ stationary solution of Eq.~\eqref{eq:SpiralEv2}.

\subsection{Spiral geometry}
\label{subsec:MeasuredGeometry}

At late times, the simulated spirals are sufficiently wound up to allow us to trace multiple spiral windings and thus classify the spiral geometry.
To do so, we focus on the entropy maps, in which the discontinuity transition is more pronounced than in temperature or density.
By locating the maximal entropy in each radial bin in the spiral plane, one can trace out the spiral and fit its pattern.

\begin{bfigure*}
\begin{center}
\myfig{
\includegraphics[height=5.2cm]{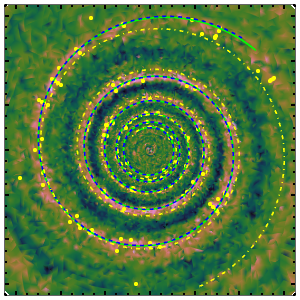}
\includegraphics[height=5.2cm]{\figeps{SpFit}}\\
\includegraphics[height=5.2cm]{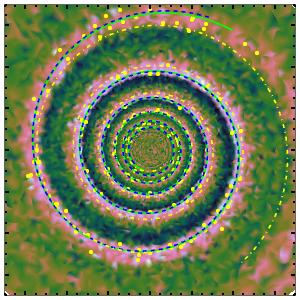}
\includegraphics[height=5.2cm]{\figeps{Sp3DFit}}\\
\includegraphics[height=5.2cm]{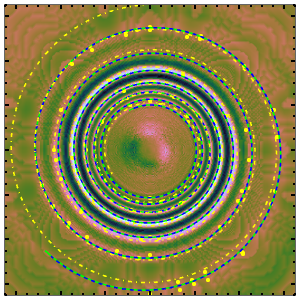}
\includegraphics[height=5.2cm]{\figeps{SpAtFit}}\\
\includegraphics[height=5.2cm]{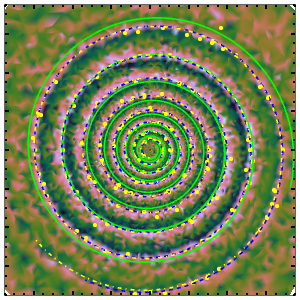}
\includegraphics[height=5.2cm]{\figeps{SpMer3DFit}}\\
}
\end{center}
\caption{
Fitting the spiral pattern in nominal runs of high resolution in offset 2D-GADGET (top row, in a $1.3\Mpc$ box), 3D-GADGET (second row; $1.4\Mpc$), and fixed-viscosity Athena (third row, $1.3\Mpc$) runs, and in the nominal 3D merger run (fourth row; $1.5\Mpc$).
The left column shows the $t=50\Gyr$ entropy in a $1\Mpc$ box, with yellow dots at the maximal entropy in each radial bin.
The right column fits these points as different spiral profiles (see legend): power law (abbrev. P; Eq.~\eqref{eq:PwrSpPattern}; solid green curve), hyperbolic (H; Eq.~\eqref{eq:HernqSpPattern}; dashed blue), or logarithmic (L; Eq.~\eqref{eq:LogSpPattern}; dot-dashed).
}
\label{fig:SpiralFits}
\end{bfigure*}

We examine three different spiral fit functions for $\phi_{d}(r)$:
hyperbolic,
\begin{equation} \label{eq:HernqSpPattern}
\phi_H(r)=\phi_0+\frac{r_0}{r+a_0}
\end{equation}
with free parameters $\phi_0$, $r_0$, and $a_0$;
logarithmic,
\begin{equation}\label{eq:LogSpPattern}
\phi_L(r)=\phi_0+\gamma^{-1}\log(r+r_0)
\end{equation}
with free parameters $\phi_0$, $r_0$, and $\gamma$;
and a general power-law,
\begin{equation}\label{eq:PwrSpPattern}
\phi_P(r)=\phi_0+\left(\frac{r_0}{r}\right)^{\lambda}
\end{equation}
with free parameters $\phi_0$, $r_0$, and $\lambda$.
Each of these functions is chosen with three free parameters, so they all have a similar fitting power.
Recall that, as shown in \S\ref{subsec:CDGeometry}, for the present, Hernquist profile, a self-similar spiral should be hyperbolic, with $a_0$ approximately given by $a$.

The spiral curves fitted for different simulations are demonstrated in Fig.~\ref{fig:SpiralFits}.
The left panels show the late-time entropy in the spiral plane, along with the best-fit spiral curves that approximately trace the maximal normalised entropy.
The right panels show the resulting best-fitting $\phi(r)$ for the above three types of spirals, and specify the chi-squared per degree of freedom ($\chi^2/\nu$) of each fit.
In all cases, we find that the hyperbolic profile provides a very good fit to the spiral, with $\chi^2/\nu<0.5$.
In offset simulations, the hyperbolic (with $a_0=30\mbox{--}60\kpc$) and power-law functions (with $\lambda\sim 0.6\mbox{--}0.8$) fit the spiral equally well, whereas a logarithmic spiral fails to provide a good fit.
In merger simulations, the hyperbolic (with $a_0=480\pm 10\kpc$) and logarithmic (with $\gamma=28.5\pm0.5$ and $r_0\simeq 100\kpc$) functions fit the spiral equally well, while a power-law spiral does not provide a good fit.

As anticipated, the structure of the discontinuity perpendicular to the spiral plane is approximately a series of nested semicircles.
In order to highlight this pattern, a few concentric circles are superimposed (as dotted cyan curves) on the perpendicular density distribution shown in the top-left panel of Fig.~\ref{fig:SSS_perp}.
At early times, the semicircles are deformed and tend to be more elongated along the $z$ axis, giving the structure a prolate appearance, as expected for a spiral structure that harbours radial flows \citep{Keshet2012}.

\subsection{Azimuthal thermal structure}
\label{subsec:SpiralAzimuthals}

\begin{figure}
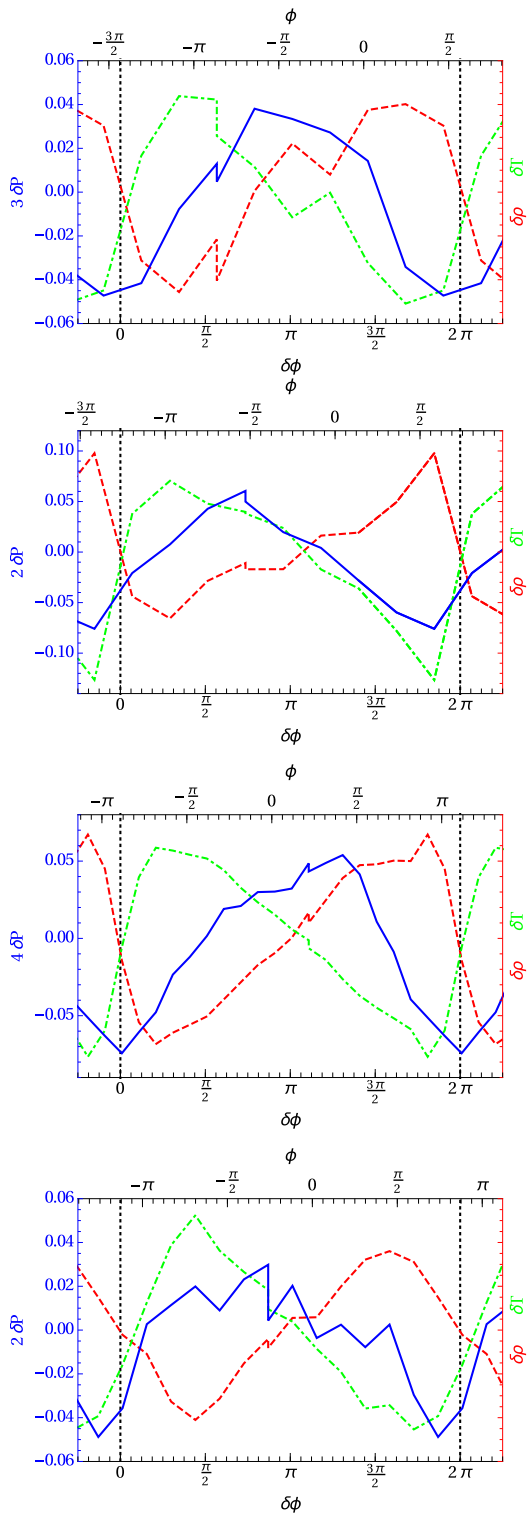

\begin{center}
\myfig{
\includegraphics[width=6.8cm]{\figeps{Prof2DT20R300}}
\vspace{0.2cm}
\includegraphics[width=6.8cm]{\figeps{Prof3DT20R400}}
\vspace{0.2cm}
\includegraphics[width=6.8cm]{\figeps{ProfAthRevNewhiT20R350}}
\vspace{0.2cm}
\includegraphics[width=6.8cm]{\figeps{ProfMer3DLN7T20R195}}
}
\end{center}
\caption{
Representative azimuthal profiles in the spiral plane ($\theta=\pi/2$), sampled at $t=20\Gyr$ (top to bottom): for offset simulations in 2D-GADGET ($r\simeq 300\kpc$), 3D-GADGET ($400\kpc$), and fixed-viscosity Athena ($350\kpc$) and for a merger simulation in 3D-GADGET ($200\kpc$).
Notations are as in Fig.~\ref{fig:SSParabola}, but using $\delta A$ instead of $\delta \tilde{A}$, \ie without normalising the azimuthal deviations by $(q+1)/(q-1)$.
Here, the precise discontinuity angle $\phi_d$ is defined by $\delta\rho(\phi_d)=0$.
}
\label{fig:SimAzimuthals}
\end{figure}

We find that in relaxed regions, the azimuthal profiles are qualitatively consistent with those derived for a quasi-steady state spiral in \S\ref{subsec:PSpiral}.
Figure \ref{fig:SimAzimuthals} illustrates the azimuthal profiles of density, temperature, and pressure in the spiral plane, for different simulations.
The figure presents the fractional deviation $\delta A$ of each quantity from its azimuthal average, as defined in Eqs.~\eqref{eq:deltaADef}--\eqref{eq:AzAvgA}.

As the figure shows, the azimuthal profiles of density and temperature are approximately linear in $\phi$, with sharp transitions on the order of $10\%$ across the discontinuity.
As expected, the density drops and the temperature jumps as one crosses outside the discontinuity, equivalent here to $\phi$ increasing above $\phi_d$, so the transition is Rayleigh-Taylor stable.

The pressure distribution is more isotropic, so the $\delta P(\phi)$ profile is multiplied in the figure, for visibility, by an estimated tightness parameter $\xi$.
The $P(\phi)$ profile is seen to be concave, with a minimum at the tangential discontinuity, and is qualitatively similar to the concave parabola anticipated for a linear density profile.
While subtle, the corresponding pressure spiral is robust, with a local minimum tracing the discontinuity.

\subsection{Radial thermal structure}
\label{subsec:SpiralRadials}

Although a spiral structure is superimposed upon the radial ICM distribution, the latter can be recovered through azimuthal averaging.
It is interesting to examine how this radial distribution differs from the initial, hydrostatic steady state.
Initially, a simulated cluster is significantly perturbed from its Hernquist steady-state only at small radii; for offset simulations, such initial deviations are confined to $r\lesssim100\kpc$.
However, radial flows, driven at early times directly by the perturbation and at later times by the emerging spiral flows, can alter the azimuthally-averaged distribution also at much larger radii.
In particular, as the spiral forms, an inflow (outflow) develops beneath (above) the discontinuity, and dissipates gradually. Recall that such a pattern, seen also in previous simulations, is inconsistent with observed flows, indicating physical processes not included in the simulations.
Figure \ref{fig:SimRadials} shows the late-time radial thermal profiles normalised to their initial, Hernquist-distribution counterparts; Figs. \ref{fig:EntropySector1}--\ref{fig:EntDx} focus on the entropy profile.

\begin{bfigure*}
\begin{center}
\myfig{
\begin{tikzpicture}
    \node[anchor=south west,inner sep=0] (image) at (0,0,0) {
\includegraphics[height=5.cm]{\figeps{NofR_SPH_RevA_t40Raw}}
\includegraphics[height=5.cm]{\figeps{NofR_SPH_RevA_t40Norm}}
    };
    \begin{scope}[x={(image.south east)},y={(image.north west)}]
        \draw[solid] (0.007,0.46) -- +(0,0.03)node[anchor=west] {};
        \draw[solid] (0.526,0.515) -- +(0,0.03)node[anchor=west] {};
    \end{scope}
\end{tikzpicture}\\
\hspace{0.63cm}
\begin{tikzpicture}
    \node[anchor=south west,inner sep=0] (image) at (0,0,0) {
\includegraphics[height=5.cm]{\figeps{TofR_SPH_RevA_t40Raw}}
\includegraphics[height=5.cm]{\figeps{TofR_SPH_RevA_t40Norm}}
    };
    \begin{scope}[x={(image.south east)},y={(image.north west)}]
        \draw[solid] (-0.001,0.49) -- +(0,0.03)node[anchor=west] {};
        \draw[solid] (0.501,0.51) -- +(0,0.03)node[anchor=west] {};
    \end{scope}
\end{tikzpicture}\\
\hspace{-0.2cm}
\begin{tikzpicture}
    \node[anchor=south west,inner sep=0] (image) at (0,0,0) {
\includegraphics[height=5.cm]{\figeps{PofR_SPH_RevA_t40Raw}}
\includegraphics[height=5.cm]{\figeps{PofR_SPH_RevA_t40Norm}}
    };
    \begin{scope}[x={(image.south east)},y={(image.north west)}]
        \draw[solid] (0.004,0.42) -- +(0,0.03)node[anchor=west] {};
        \draw[solid] (0.527,0.52) -- +(0,0.03)node[anchor=west] {};
    \end{scope}
\end{tikzpicture}\\
\hspace{0.4cm}
\begin{tikzpicture}
    \node[anchor=south west,inner sep=0] (image) at (0,0,0) {
\includegraphics[height=5.cm]{\figeps{KofR_SPH_RevA_t40Raw}}
\includegraphics[height=5.cm]{\figeps{KofR_SPH_RevA_t40Norm}}
    };
    \begin{scope}[x={(image.south east)},y={(image.north west)}]
        \draw[solid] (0.003,0.43) -- +(0,0.03)node[anchor=west] {};
        \draw[solid] (0.51,0.51) -- +(0,0.03)node[anchor=west] {};
    \end{scope}
\end{tikzpicture}
}
\end{center}
\caption{
Evolved, $t=40\Gyr$ radial (azimuthally-averaged) profiles of electron number density (top row) and gas temperature (second row), pressure (third row), and adiabat (bottom row), shown before (left column) and after (right) normalising to the initial, Hernquist distribution (dot-dashed black in left column).
Results shown for offset simulations of 2D-GADGET (solid green), 3D-GADGET (dashed blue and cyan), and  fixed-viscosity Athena (dot-dashed red and orange); the latter two shown both within (short dashing) and perpendicular (long dashing and lighter colour) to the spiral plane.
Azimuthal volume averaging is used; mass-averaging gives qualitatively similar results.
The Athena simulations show a constant-entropy core and a linear entropy region (dotted black curves in bottom-right panel); the flat entropy profile in the core is gradually replaced by the linear profile as resolution is raised.
}
\label{fig:SimRadials}
\end{bfigure*}

As expected, the radial flows mix the plasma, pushing cold, dense, low-entropy gas outwards and bringing hot, dilute, high-entropy gas inwards.
This mixing is dominated by the initial perturbation, whereas the subsequent spiral flows in such simulations modify the configuration in the opposite direction, with the low-entropy gas inside the discontinuity flowing inwards \citep{Keshet2012}.
Consequently, at late times we find that the entropy inside $r\sim 1\Mpc$ increases with respect to its initial value.
This increase is associated both with a decline in density and a more modest rise in temperature, such that the pressure decreases.
As the figure demonstrates, the thermal distributions inside and perpendicular to the spiral plane are very similar to each other.
The temperature profile is found to be similar also among different simulation codes, whereas the lowered density does differ somewhat among codes, corresponding to different efficiencies of gas ejection from the core.

The Athena simulations show a somewhat different behaviour near the centre, with the temperature rising and density dropping rapidly as $r$ decreases, giving a constant entropy core.
This strongly-mixed core, generated by a convective instability driven by the initial perturbation and subsequent flows, becomes larger for a stronger perturbation or for weaker viscosity.
The core converges over a $\lesssim 10\Gyr$ timescale, washing out the spiral structure near the centre.
Although the core is substantially larger than the resolution-induced mixing region seen in our nominal set-up in the absence of a perturbation, we find that the core progressively shrinks as the resolution increases; this effect is particularly strong for merger simulations, as shown in Fig. \ref{fig:EntRes}.

The different behaviour of GADGET simulations appears to be better converged with resolution, at least for $r\gtrsim 20\kpc$, but these simulations are known to poorly resolve the Kelvin-Helmholtz and convective instabilities \citep{Mitchell09, ValkeEtAl10, Tricco19} responsible for the Athena cores.
Hence, we focus on the Athena simulations, as a better representation of the ICM behaviour in the central regions and on small scales.
One should keep in mind, however, that such simulations cannot resolve the evolution of the core under more realistic conditions, where AGN feedback, magnetic fields, heat conduction, and turbulent effects may play an important role.

\subsection{Entropy profile}
\label{subsec:SpiralEntropy}

\begin{bfigure*}
\begin{center}
\begin{tikzpicture}
    \node[anchor=south west,inner sep=0] (image) at (0,0,0) {
    \includegraphics[width=0.475\linewidth,trim={0 0 0 0},clip]{\figeps{Sec1EntSteps2}}
    \includegraphics[width=0.475\linewidth,trim={0 0 0 0},clip]{\figeps{MerM10EntSteps2}}
    };
    \begin{scope}[x={(image.south east)},y={(image.north west)}]
    \end{scope}
\end{tikzpicture}
\end{center}
\caption{
The adiabat profile along a radial ray in the spiral plane.
Left: Late-time offset simulations (notations as in Fig.~\ref{fig:SimRadials}; 3D-GADGET and Athena profiles are manually offset to higher $K$, for visibility), showing $K(r)$ both azimuthally averaged (thin curves) and in a $20^\circ$ sector (thick curves).
Right: Athena merger simulation with $r_M=10$, showing $K(r)$ in a $20^\circ$ sector at $t=\{5,10,20,40\}\Gyr$ (red to blue solid curves; earlier-time profiles increasingly offset to lower $K$, for visibility).
\label{fig:EntropySector1}
}
\end{bfigure*}

The entropy profile is particularly illuminating, because in our adiabatic simulations, entropy directly traces the mixing of the gas.
Figure \ref{fig:EntropySector1} presents the profile $K(r)$ along a radial ray in the spiral plane, showing the sequential jumps in the adiabat at each crossing of the spiral discontinuity.
As the figure shows, offset (left panel) and merger (right) Athena simulations typically generate spirals sufficiently tight and discontinuity jumps sufficiently high to flatten out the local radial entropy profile between discontinuities, and even produce locally negative $\partial_r K$.
Such a profile becomes convectively unstable locally; in a self-similar spiral, this implies a violation of the stability condition (\ref{eq:LocalConvectionInstability}).
Under such conditions, the $K(r)$ profile in a given sector should saturate onto a series of flat ($\partial_r K=0$) steps, as indeed seen in the figure.

Furthermore, once a convective instability emerges locally, between adjacent spiral windings, the global entropy profile should not remain too shallow, as discussed in \S\ref{subsec:RadFlowAndInst}.
Namely, under the assumptions leading to Eq.~(\ref{eq:GlobalConvectiveInstability}), the azimuthally-averaged $\bar{K}(r)$ profile outside the convective core should not remain sublinear.
As a result, one expects $\bar{K}(r)$ outside the convective core to relax into a linear, $\bar{K}\propto r$ profile, bridging between the $\bar{K}(r)=\const$ centre and the superlinear $\bar{K}(r)$ periphery.
Such a behaviour is indeed seen in our Athena simulations, as hinted by the dotted black curves in the bottom right panel of Fig.~\ref{fig:SimRadials}.
The effect is shown in more detail in Figs.~\ref{fig:EntEv}--\ref{fig:EntDx}, which depict the $\bar{K}/r$ profile to highlight $\bar{K}/r\simeq \const$ regions in both offset (left panels) and merger (right panels) simulations.

\begin{bfigure*}
\begin{center}
\begin{tikzpicture}
    \node[anchor=south west,inner sep=0] (image) at (0,0,0) {
\myfig{
\includegraphics[width=0.475\linewidth]{\figeps{KperR_Athena_FixedMu_Ev}}
\includegraphics[width=0.475\linewidth]{\figeps{KperR_Athena_Mer_rM10_Ev}}
}
    };
    \begin{scope}[x={(image.south east)},y={(image.north west)}]
        \draw[solid] (0.003,0.34) -- +(0,0.03)node[anchor=west] {};
        \draw[solid] (0.503,0.35) -- +(0,0.03)node[anchor=west] {};
    \end{scope}
\end{tikzpicture}
\end{center}
\caption{
Temporal evolution of the azimuthally-averaged $\bar{K}/r$ profile in high-resolution fixed-viscosity offset (left panel) and $r_M=10$ merger (right panel) Athena simulations at times $t=5$, $10$, $20$, $30$, $40$, and $50$ Gyr (short pink to long blue dashing). Also shown are the initial, Hernquist profile (dot-dashed black) and the observed profile \citep[dotted purple with shaded region for the dispersion in normalization among different systems, from][]{ReissKeshet2015}.
}
\label{fig:EntEv}
\end{bfigure*}

\begin{bfigure*}
\begin{center}
\begin{tikzpicture}
    \node[anchor=south west,inner sep=0] (image) at (0,0,0) {
    \myfig{
\includegraphics[width=0.475\linewidth]{\figeps{KperR_Athena_FixedMu_t40_Res}}
\includegraphics[width=0.475\linewidth]{\figeps{KperR_Athena_Merger_Res_rM10_Dashing}}
}
    };
    \begin{scope}[x={(image.south east)},y={(image.north west)}]
        \draw[solid] (0.003,0.35) -- +(0,0.03)node[anchor=west] {};
        \draw[solid] (0.503,0.35) -- +(0,0.03)node[anchor=west] {};
    \end{scope}
\end{tikzpicture}
\end{center}
\caption{
Same as Fig.~\ref{fig:EntEv} at $t=40\Gyr$ for fixed viscosity simulations of low (red), medium (orange), nominal (green) and high (blue) resolutions (short to long dashing).
}
\label{fig:EntRes}
\end{bfigure*}

\begin{figure*}
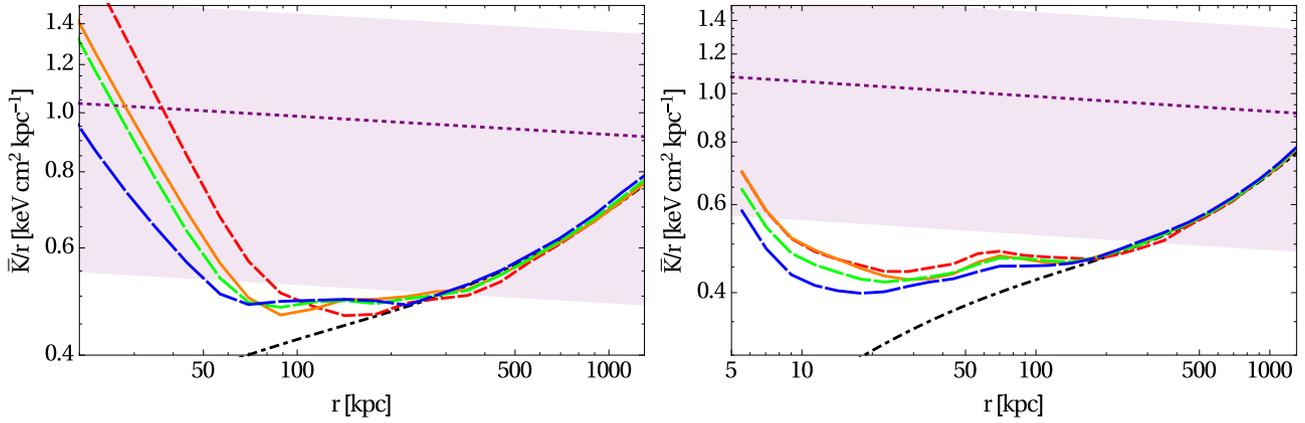

\begin{center}
\begin{tikzpicture}
    \node[anchor=south west,inner sep=0] (image) at (0,0,0) {
\myfig{
\includegraphics[width=0.475\linewidth]{\figeps{KperR_Athena_FixedMu_NSpitzC}}
\includegraphics[width=0.475\linewidth]{\figeps{KperR_Athena_Merger_Visc_rM10}}
}
    };
    \begin{scope}[x={(image.south east)},y={(image.north west)}]
        \draw[solid] (0.003,0.35) -- +(0,0.03)node[anchor=west] {};
        \draw[solid] (0.503,0.35) -- +(0,0.03)node[anchor=west] {};
    \end{scope}
\end{tikzpicture}
\end{center}
\caption{
Same as Fig.~\ref{fig:EntEv} for nominal resolution with Spitzer viscosity (solid orange curve) or fixed $\mu/\mu_S(3\keV)=1$, $3$, and $10$ (short red to long blue dashing).
}
\label{fig:EntNSpitz}
\end{figure*}

\begin{bfigure*}
\begin{center}
\begin{tikzpicture}
    \node[anchor=south west,inner sep=0] (image) at (0,0,0) {
\myfig{
\includegraphics[width=0.475\linewidth]{\figeps{KperR_Athena_FixedMu_dx}}
\includegraphics[width=0.475\linewidth]{\figeps{KperR_Athena_Merger_Res_rM_HR_UHR}}
}
    };
    \begin{scope}[x={(image.south east)},y={(image.north west)}]
        \draw[solid] (0.003,0.35) -- +(0,0.03)node[anchor=west] {};
        \draw[solid] (0.503,0.35) -- +(0,0.03)node[anchor=west] {};
    \end{scope}
\end{tikzpicture}
\end{center}
\caption{
Same as Fig.~\ref{fig:EntEv} for weak to strong perturbations (short red to long blue dashing).
Offset simulations (left panel, nominal resolution) with initial $dx/a=0.05$, $0.1$, $0.2$, $0.3$, and $0.5$.
Merger simulations (right panel, high resolution) with mass ratios $r_M=20$, $10$, $5$, and $3$; the latter two results are poorly converged, as demonstrated by the flattening of the $r_M=5$ bump at $r\simeq 30\kpc$ at an even higher resolution (dotted green).
}
\label{fig:EntDx}
\end{bfigure*}

Indeed, a linear, $\bar{K}(r)\propto r$ region emerges in our simulations quite robustly.
In early stages, this linear region shows oscillations (Fig.~\ref{fig:EntEv}) indicative of an instability, gradually dissipating due to radial flows.
The linear region, spanning about an order of magnitude in $r$ in our high-resolution simulations, becomes more extended and less interrupted by oscillations at later times (Fig.~\ref{fig:EntEv}), higher resolutions (Fig.~\ref{fig:EntRes}), or stronger viscosities (Fig.~\ref{fig:EntNSpitz}).
While the normalization $\bar{K}/r\simeq\const$ in our simulations is smaller than that of the observed universal profile (dotted purple line in these figures), the former increases for stronger perturbations (Fig.~\ref{fig:EntDx}), \ie as a larger volume of the cluster is affected by the spiral structure.

The $\bar{K}(r)$ profile remains linear in regions that contained a tight or sharp spiral even after it dissipated, as seen by comparing the right panels of Figs.~\ref{fig:EntropySector1} and \ref{fig:EntEv}.
Therefore, a spiral perturbation traversing a large ICM volume can imprint a linear $\bar{K}(r)$ profile even if no persistent spiral is observed.
In addition, feedback and heating processes not included in our simulations could sustain the spiral over larger scales, extend the linear $\bar{K}(r)$ region, and further raise its $\bar{K}/r$ normalization.

\section{Summary and discussion}
\label{sec:Discussion}

We show, analytically and numerically, how a combination of ICM mixing and rotation naturally leads to the robust formation of a thermal quasi-spiral structure over a few Gyr timescales, and quantify the resulting quasi-steady state at late times.
Thus, a wide range of perturbations, strong enough to create a tangential discontinuity and deposit sufficient angular momentum, lead to a similar structure (see Fig.~\ref{fig:DifferentEvolution}--\ref{fig:AthenaViscosity}), with viscosity regulating the evolution and tightness of the spiral through the dissipation of differential rotation (Eq.~\eqref{eq:CFPattern} and Figs.~\ref{fig:omegaVSr}, \ref{fig:omegaVSt}).

The early-time evolution is sensitive to the underlying plasma distribution, the details of energy and angular-momentum deposition, the number of dimensions, viscosity, and additional physical processes such as feedback and radiative cooling.
However, the late-time outcome is an approximately self-similar pattern (Eqs.~\eqref{eq:SSSAnsatz}, \eqref{eq:LinearDRho}, \eqref{eq:LinearDP}), fixed mainly by the gravitational potential (Eq.~\eqref{eq:SSSpiral}), the initial density distribution, and the dimensionless tightness $\xi$ and contrast $q$ parameters, which are related to each other; see \S\ref{subsec:SelfSimilar}.
While the self-similar nature of the flow becomes increasingly transparent at late times, as the pattern approaches uniform slow rotation, its basic attributes already manifest a few Gyr after the perturbation.
We verify the results using a range of Eulerian and Lagrangian simulations of 2D and 3D, merger and offset, clusters.

The late-time structures emerging in these simulations are robust, similar to each other, and of morphological and thermal features that agree with observations, suggesting that a simple quasi-steady state can be used as a basis for modelling the ICM even when additional physical processes, in particular feedback and radiative cooling, are incorporated.
Similar late-times structures are obtained in 2D and in the equatorial plane in 3D, indicating that radial flows are not essential and that the plane can be analysed in isolation.
While the evolution of the quasi-spiral structure and its ultimate tightness depend on viscosity, we obtain qualitatively similar structures for different $\mu(T)$ prescriptions, facilitating simplified models based on a uniform viscosity (see \S\ref{subsec:PlanarEv}).

The quasi-spiral structure is strongly constrained by the presence of a $\phi_d(r,\theta)\propto \Phi(r)$ discontinuity manifold, combining a trailing spiral in the equatorial plane with concentric semicircles perpendicular to the plane, in resemblance of a snail shell (see Fig.~\ref{fig:setup}).
In particular, the hyperbolic spiral pattern corresponding to a Hernquist potential is recovered for a wide range of offset and merger simulations (see Fig.~\ref{fig:SpiralFits}).
The tightness of the spiral, given by $\xi$ (see Eq.~\eqref{eq:NSr2}), increases in time until differential rotation is dissipated.
Once the tangential discontinuity has been created, it remains embedded in the ICM and cannot disappear, although its contrast may dissipate without feedback or other processes that sustain it.

The thermal structure is determined by the discontinuity density contrast $q$, which is approximately constant close to the spiral plane; a generalization for the full 3D structure is outlined in \S\ref{subsubsec:SelfSimilarAnsatz} and followed-up in Ghosh et al. (in preparation).
A subtle pressure spiral must emerge, both before (\S\ref{subsec:NonSelfSimilar}) and after (\S\ref{subsec:SelfSimilar}) the structure relaxes to its self-similar state, in order to entrain the hot and cold plasma phases at the same radii, preventing them from buoyantly rising or sinking.
Thus, while an evolved spiral tends toward linear $\rho(\phi)$ and $T(\phi)$ profiles (at a given radius $r$), the pressure $P(\phi)$ given by their product is not constant, but rather a concave parabola minimized as $\phi=\phi_d$. This result agrees with theory (Eqs.~\eqref{eq:LinearDRho}--\eqref{eq:deltaPTight} and Figs.~\ref{fig:SSParabola} and \ref{fig:PParabola}), numerical simulations (Fig.~\ref{fig:SimAzimuthals}), and observations.

The radial, azimuthally-averaged profiles $\bar{T}$ of temperature and $\bar{K}$ of the adiabat are typically enhanced by the presence of a quasi-spiral structure (Figs.~\ref{fig:SSTandK} and \ref{fig:SimRadials}), while the density and pressure profiles can remain unchanged by a relaxed, self-similar structure.
In a given angular sector, the radial profiles of density and entropy increasingly flatten between spiral windings of the discontinuity as $\xi$ or $q$ increase.
Extreme values of these parameters (Eqs.~\eqref{eq:LocalRTInstability}--\eqref{eq:LocalConvectionInstability}) can thus lead to convective and potentially even Rayleigh-Taylor instabilities.
Indeed, Athena simulations resolve a tightening of the spiral sufficient to induce a local convective instability within radial sectors bounded by consecutive spiral windings.
This instability leads to flat entropy steps $K(r)$ in angular sectors (Fig.~\ref{fig:EntropySector1}) and, at least at low resolution, to a flat-entropy ($K=\const$) core in the centre of the cluster.

We find that between the flat-entropy core and the steep rising-entropy periphery, the spiral structure imprints a linear, $\bar{K}(r)\propto r$ adiabat that persists after the spiral dissipates.
This linear behaviour develops in regions that show flat $K(r)$ steps in angular sectors, and involves temporary oscillations in the azimuthally-averaged $\bar{K}(r)$, suggesting some global convective instability.
The $\bar{K}(r)\propto r$ behaviour becomes more pronounced and spans a larger region at later times (Fig.~\ref{fig:EntEv}), higher resolutions (Fig.~\ref{fig:EntRes}), or stronger viscosities (Fig.~\ref{fig:EntNSpitz}).
Our numerical results can be understood (\S\ref{subsec:RadFlowAndInst}) in terms of a modified convective instability, associated with the motion of radial segments of constant $K$ that are sustained on a short timescale by the local instability.

Interestingly, a universal, linear $\bar{K}(r)$ profile is found in well-deprojected galaxy cluster observations, and there is evidence showing that this profile is regulated by a dynamical process.
Our results suggest that even transient spiral structures could be responsible for imprinting such a profile onto the ICM.
Unlike our simulations, the $\bar{K}(r)$ profile in observations shows a linear behaviour spanning the entire cluster, resulting in a universal normalization; reproducing such a profile would require stronger perturbations (see Fig.~\ref{fig:EntDx}), repeated perturbations, or additional physical processes.

While the quasi-spiral solution can serve as a basis for modelling the spiral thermal structures typical of the ICM, it lacks the nearly sonic outflows inside CFs, the slow inflows outside CFs, radiative cooling, radio bubbles, deviations from hydrostatic equilibrium associated with magnetic field layers, and other properties of observed clusters. Outflows, inflows, and the magnetic layers they induce, while not essential for the quasi-steady state, can be incorporated in the model, probably accelerating the spiral evolution and protecting the cool core from the cooling instability.
Finally, as we obtain similar structures for a wide range of perturbations, and no present model or simulation reproduces all of the essential features outlined above, specific merger scenarios invoked in the literature to explain a given spiral ICM observation may be non-unique.

\MyMNRAS{\section*{Acknowledgements}}
We thank I. Gurwich, Y. Gal, Y. Moyal, E. Malka, I. Reiss, and Y. Lyubarsky for helpful discussions.
This research was supported by the Israel Science Foundation (Grants No. 1769/15 and 2126/22), by the IAEC-UPBC joint research foundation (Grant No. 300/18), and by the Ministry of Science, Technology \& Space, Israel, and has received funding from the GIF (Grant No. I-1362-303.7/2016).

\section*{Data Availability} The data generated from computations are reported in the paper, and any additional data will be made available upon reasonable request to the corresponding author.

\bibliography{Clusters}

\MyMNRAS{\label{lastpage}}

\appendix

\section{Pressure spiral derivation}
\label{sec:AppPSpiral}

The spiral discontinuity can be incorporated in the integral in Eq.~\eqref{eq:PSol} through piecewise integration or using identities such as $(\phi\mbox{ mod }2\pi)=\pi-2\sum_{k=1}^\infty \sin(k \phi)/k$.
For a linear $\rho(\phi)$ profile, suffice to collect the contributions to $\partial_{\phi\phi}P$ from discontinuities crossed along a radial ray, as shown in Eq.~\eqref{eq:Pphiphi}.

For instance, for a steady-state $P_0\propto r^\lambda$ profile perturbed by a $\phi=2\pi c\, r^{\lambda_c}$ spiral pattern with the linear azimuthal density profile \eqref{eq:LinearDRho}, the pressure distribution derived from Eq.~\eqref{eq:PSol} is then given by
\begin{align}
& \frac{q+1}{q-1}\left(\frac{P}{P_0}-1\right) = 1+2j+2\tilde{\phi} - \frac{2 r^{\lambda_c} c }{1+\lambda_c/\lambda}  \\
& \quad \quad \quad \quad + 2\frac{\zeta\left(-\frac{\lambda}{\lambda_c},\tilde{\phi}+j\right) - \zeta\left(-\frac{\lambda}{\lambda_c},\tilde{\phi}\right)-(\tilde{\phi}+j)^{\lambda/\lambda_c}}{r^\lambda c^{\lambda/\lambda_c}} \coma \nonumber
\end{align}
where $\tilde{\phi}\equiv \phi/(2\pi)$, $j\equiv \lfloor c\, r^{\lambda_c}-\tilde{\phi} \rfloor$, $\zeta$ is the Hurwitz zeta function, and the free parameters $\lambda$, $c$, and $\lambda_c$ are assumed constant.
Similar but more lengthy expressions can be derived for more sophisticated spiral patterns, density profiles, contrast scalings, and ICM distributions, including the hyperbolic spiral in a Hernquist ICM shown in Fig.~\ref{fig:PParabola}.

\end{document}